%% file: epic_arxiv_v4.tex
\documentclass{emulateapj}

\usepackage{graphicx} 
\usepackage{longtable} 
\usepackage{rotating}
\usepackage{tabularx}
\usepackage{hyperref}
\usepackage{easylist}
\usepackage{amsmath}
\usepackage{bm}
\usepackage{array} 
\usepackage{amssymb,amsmath} 

\newcommand{\numax}{\mbox{$\nu_{\rm max}$}}

\newcommand{\kep}{\mbox{\textit{Kepler}}}
\newcommand{\teff}{\mbox{$T_{\rm eff}$}}
\newcommand{\logg}{\mbox{$\log g$}}
\newcommand{\feh}{\mbox{$\rm{[Fe/H]}$}}
\newcommand{\msun}{\mbox{$M_{\sun}$}}
\newcommand{\rsun}{\mbox{$R_{\sun}$}}
\newcommand{\rhosun}{\mbox{$\rho_{\sun}$}}
\newcommand{\rearth}{\mbox{$R_{\ensuremath{\oplus}}$}}

\newcommand{\nkepstars}{\mbox{17,068}}

\newcommand{\ntotkp}{\mbox{156,456}}

\newcommand{\nclass}{\mbox{138,600}}

\newcommand{\nkmdwarfs}{\mbox{41}}

\newcommand{\nfgdwarfs}{\mbox{36}}

\newcommand{\nkgiants}{\mbox{21}}

\include{journals}

\slugcomment{accepted for publication in ApJS}

\shorttitle{The K2 Ecliptic Plane Input Catalog}
\shortauthors{Huber et al.}

\begin{document}

\title{The K2 Ecliptic Plane Input Catalog (EPIC) and Stellar Classifications of \\
138,600 Targets in Campaigns 1--8}

\author{
Daniel Huber\altaffilmark{1,2,3}, 
Stephen T.\ Bryson\altaffilmark{4}, 
Michael R.\ Haas\altaffilmark{4}, 
Thomas Barclay\altaffilmark{4,5},
Geert Barentsen\altaffilmark{4,5}, \\
Steve B.\ Howell\altaffilmark{4},
Sanjib Sharma\altaffilmark{1}, 
Dennis Stello\altaffilmark{1,3}, and
Susan E.\ Thompson\altaffilmark{2,4}
}

\altaffiltext{1}{Sydney Institute for Astronomy (SIfA), School of Physics, University of 
Sydney, NSW 2006, Australia; daniel.huber@sydney.edu.au}
\altaffiltext{2}{SETI Institute, 189 Bernardo Avenue, Mountain View, CA 94043, USA}
\altaffiltext{3}{Stellar Astrophysics Centre, Department of Physics and Astronomy, Aarhus 
University, Ny Munkegade 120, DK-8000 Aarhus C, Denmark}
\altaffiltext{4}{NASA Ames Research Center, Moffett Field, CA 94035, USA}
\altaffiltext{5}{Bay Area Environmental Research Inst., 560 Third St., West Sonoma, 
CA 95476, USA}

\begin{abstract}
The K2 Mission uses the \kep\ spacecraft to obtain high-precision photometry 
over $\approx$\,80 day campaigns in the ecliptic plane. The Ecliptic 
Plane Input Catalog (EPIC) provides coordinates, photometry and kinematics based on a 
federation of all-sky catalogs to support target selection and target management for the 
K2 mission. We describe the construction of the EPIC, as well as modifications and 
shortcomings of the catalog. \kep\ magnitudes (\textit{Kp}) are 
shown to be accurate to $\approx 0.1$\,mag for the \kep\ field, and the EPIC is typically 
complete to $Kp\approx17$ ($Kp\approx19$ for campaigns covered by SDSS). 
We furthermore classify \nclass\ targets in Campaigns 1--8 ($\approx$\,88\% of the full 
target sample) using colors, proper motions, spectroscopy, parallaxes, and 
galactic population synthesis models, with typical uncertainties for G-type stars of 
$\approx$\,3\% in \teff, $\approx$\,0.3\,dex in \logg, $\approx$\,40\% in radius, 
$\approx$\,10\% in mass, and $\approx$\,40\% in distance. Our results show that stars 
targeted by K2 are dominated by K--M dwarfs ($\approx$\,\nkmdwarfs\% of all selected targets), 
F--G dwarfs ($\approx$\,\nfgdwarfs\%) 
and K giants ($\approx$\,\nkgiants\%), consistent with key K2 science programs to search for 
transiting exoplanets and galactic archeology studies using oscillating red giants. 
However, we find a significant variation of the fraction of cool dwarfs with 
galactic latitude, indicating a target selection bias due to interstellar reddening 
and the increased contamination by giant stars near the galactic plane.
We discuss possible systematic errors in the derived stellar properties, 
and differences to published classifications for 
K2 exoplanet host stars. The EPIC is hosted at the Mikulski Archive 
for Space Telescopes (MAST): \url{http://archive.stsci.edu/k2/epic/search.php}.
\end{abstract}



\section{Introduction}
The NASA \kep\ Mission \citep{borucki10,koch10b} delivered breakthrough discoveries in 
exoplanet science and stellar astrophysics by obtaining high precision photometry of 
a single field for four years. To facilitate target selection 
and to enable the definition of optimal 
apertures for the extraction of light curves, a dedicated ground-based imaging 
campaign of the \kep\ field was federated with available all-sky catalogs to create the 
Kepler Input Catalog \citep[KIC,][]{brown11}, which contains $\approx$13 million 
sources centered on the \kep\ field. In addition to broadband photometry 
the KIC includes \kep\ magnitudes (\textit{Kp}), which were defined using $gri$ magnitudes 
through a calibration of the expected flux in the \kep\ bandpass, and 
stellar properties derived from fitting photometry to synthetic colors from 
model atmospheres. The KIC has been an indispensable tool for a variety of \kep\ science 
such as target selection \citep{batalha10}, 
planet-candidate catalogs \citep{borucki11b,borucki11,batalha12},
studies of planet occurrence rates \citep{youdin11,howard11,fressin13}, and the 
derivation of fundamental properties of oscillating stars \citep{kallinger10,chaplin11a}.

Following the failure of the second of four reaction wheels, the \kep\ mission 
ended in May 2013 due to the inability to maintain fine-point in the original 
field. The spacecraft was subsequently repurposed as the K2 mission to 
observe fields along the ecliptic plane using the 
solar pressure to maintain fine-point along the roll axis of the spacecraft \citep{howell14}. 
Methodologies to account for systematic 
variability caused by the spacecraft thruster firings are continuously being developed 
\citep{vanderburg14,aigrain15,angus15,huang15,lund15} and have demonstrated 
that the photometric precision of K2 is at least within a factor 
of 2 of \kep.
Early K2 science results include the discovery of exoplanets around bright GKM dwarfs 
\citep{crossfield15,vanderburg15,dfm15,petigura15}, eclipsing binary 
stars \citep{armstrong15,laCourse15},
as well as the detection of pulsations across the HR diagram such as 
subgiants \citep{chaplin15}, red giants \citep{stello15}, white dwarfs \citep{hermes14},  
RR Lyrae stars \citep{molnar15,kurtz16}, O stars \citep{buysschaert15} 
and subdwarf B stars \citep{jeffery14}.

In this paper we describe the Ecliptic Plane Input Catalog (EPIC) for the K2 mission, 
which serves a similar role as the KIC for the \kep\ mission. 
The primary purpose of the catalog is to provide positions and \kep\ magnitudes for 
target management and the definition of optimal apertures.  A secondary goal is to provide observables 
to facilitate target selection by the community. The EPIC does not contain stellar properties for 
all sources, but in this work we perform classifications of \nclass\ targets which were 
selected for observations to support community investigations of K2 data.
Since K2 field positions are not fixed for the duration of the mission, the EPIC is 
continuously updated to include new campaign fields. 
At the time of writing of this 
paper the EPIC has been released for Campaigns 1--13 (C1--C13), 
with no substantial changes expected for future fields.

\section{Catalog Construction}

\subsection{Input Sources and Cross-Matching}
\label{sec:xmatching}

EPIC sources are drawn from several publicly available all-sky catalogs:
the Hipparcos catalog \citep{vanleeuwen07b}, the Tycho-2 
catalog \citep{hog00}, the fourth US Naval Observatory CCD Astrograph 
Catalog \citep[UCAC4,][]{zacharias13}, the Two Micron All Sky 
Survey \citep[2MASS,][]{skrutskie06}, and data release 9 of the Sloan Digital Sky 
Survey \citep[SDSS DR9,][]{ahn12}.  Each catalog was downloaded from 
Vizier\footnote{\url{http://webviz.u-strasbg.fr/viz-bin/VizieR}} using a search radius of 9 degrees 
centered on each K2 campaign field, which is large enough to allow for potential small 
adjustments to the field of view after the EPIC is made available to the 
community.

The quality of the photometry in the input catalogs varies significantly. To ensure 
some homogeneity, the following quality cuts and transformations 
were applied to the observables:

\begin{list}{\labelitemi}{\leftmargin=1em}
\item Tycho-2: $B_{T}$ and $V_{T}$ photometry were converted into the Johnson $BV$ by 
interpolating Table 2 in \citet{bessell00}.

\item UCAC4: $g'r'i'$ photometry (which originates from the AAVSO Photometric All-Sky 
Survey, APASS\footnote{\url{http://www.aavso.org/apass}}) 
was converted to the Sloan $gri$ system using the 
transformation equations given on the SDSS 
website\footnote{\url{http://www.sdss.org/dr7/algorithms/jpeg_photometric_eq_dr1.html}}.  
APASS $BVgri$ uncertainties which are zero or 0.01 mag in UCAC4 
were replaced with average 
APASS uncertainties as a function of $BVgri$ magnitude.

\item 2MASS: All sources brighter than $J = 5$ mag were discarded due to known saturation 
problems, and all sources with a $J$-band quality flag worse than C were removed. 
All $H$ and $K$-band measurements were also omitted for quality flags 
worse than C.

\item SDSS DR9: Only targets with clean photometry, $r$ magnitudes lower than 20, and 
photometry errors lower than 0.5 mag were retained. 
\end{list}

\noindent
Each catalog was cross-matched for overlapping sources.  Published matches 
(Hipparcos-Tycho, UCAC-Tycho-2MASS) were adopted when available, otherwise sources 
were matched by finding the closest object within 3 arcseconds. To eliminate chance 
matches with background objects, the $V$ or $g$ magnitude of the target and matched 
source were required to agree within 1.5 mag, which is the typical maximum 3$\sigma$ 
uncertainty in a given passband. For cross-matches without common passbands 
(such as Tycho-2MASS, or 2MASS-SDSS), the $V$ and $g$ magnitudes of the matched source 
were estimated using Johnson-2MASS-SDSS transformations of \citet{bilir05} and 
\citet{bilir08}.  For transformed magnitudes, the matching criteria were 
conservatively set to 2 mag for $g$-band magnitudes estimated from $BV$, and 4 mag for 
$g$-band magnitudes estimated from $JHK$.  While this procedure should eliminate most 
erroneous cross-matches, it may result in duplicate entries.  Visual inspection of 
images from the second Palomar Observatory Sky Survey 
(POSS-II)\footnote{\url{http://archive.stsci.edu/cgi-bin/dss_form}}
showed that duplicate entries are relatively rare (see also Figure \ref{fig:dss}).

\begin{table}
\begin{small}
\begin{center}
\caption{Ecliptic Plane Input Catalog columns.}
\begin{tabular}{l l l l}        
\hline      
Name  &	Unit			     &		Description				     \\
\hline      
EPIC	     &  none			     &  	K2 Identifier				     \\
HIP	     &  	none		     &  		Hipparcos Identifier		     \\
TYC	     &  	none		     &  		Tycho-2 Identifier		     \\
UCAC	     &  none			     &  	UCAC4 Identifier			     \\
2MASS	     &  none			     &  	2MASS Identifier			     \\
SDSS	     &  none			     &  	SDSS Identifier 			     \\
Object type  &  none			     &  	Object Type Flag \\
				&					&		[STAR, EXTENDED]	     \\
Kepflag	     &  none			     &  	Kepler Magnitude Flag \\
			&						&		[gri, BV, JHK, J]      \\
RA	     	 &  	degrees 	     &  		Right Ascension (JD2000)	     \\
Dec	     	 &  	degrees 	     &  		Declination (JD2000)		     \\
pmra	     &  mas/yr	     &  	Proper Motion in RA			     \\
pmdec	     &  mas/yr	     &  	Proper Motion in DEC			     \\
plx	     	 &  	mas      &  	Parallax				     \\
Bmag	     &  magnitude		     &  	Johnson B band magnitude		     \\
Vmag	     &  magnitude		     &  	Johnson V band magnitude		     \\
umag	     &  magnitude		     &  	Sloan u band magnitude			     \\
gmag	     &  magnitude		     &  	Sloan g band magnitude			     \\
rmag	     &  magnitude		     &  	Sloan r band magnitude			     \\
imag	     &  magnitude		     &  	Sloan i band magnitude			     \\
zmag	     &  magnitude		     &  	Sloan z band magnitude			     \\
Jmag	     &  magnitude		     &  	2MASS J band magnitude			     \\
Hmag	     &  magnitude		     &  	2MASS H band magnitude			     \\
Kmag	     &  magnitude		     &  	2MASS K band magnitude			     \\
KepMag	     &  magnitude		     &  	Kepler magnitude (Kp)			     \\
\hline
NOMAD		&   none				&	NOMAD1 Identifier \\
Mflg 		&	none 				&	2MASS Flags \\
			&					    &	[Qflg-Rflg-Bflg-Cflg-Xflg-Aflg] \\
proxy 		&	as 					&	2MASS nearest neighbor \\
\hline												     
\end{tabular} 
\flushleft Notes: Bracketed items list the possible values 
for catalog flags (see text for details). Columns containing uncertainties and columns 
which are typically not populated for any given EPIC source (such as stellar properties) 
are omitted from this table for clarity.
\label{tab:epic}
\end{center}

\end{small}
\end{table}

\begin{figure*}
\begin{center}
\resizebox{8.75cm}{!}{\includegraphics{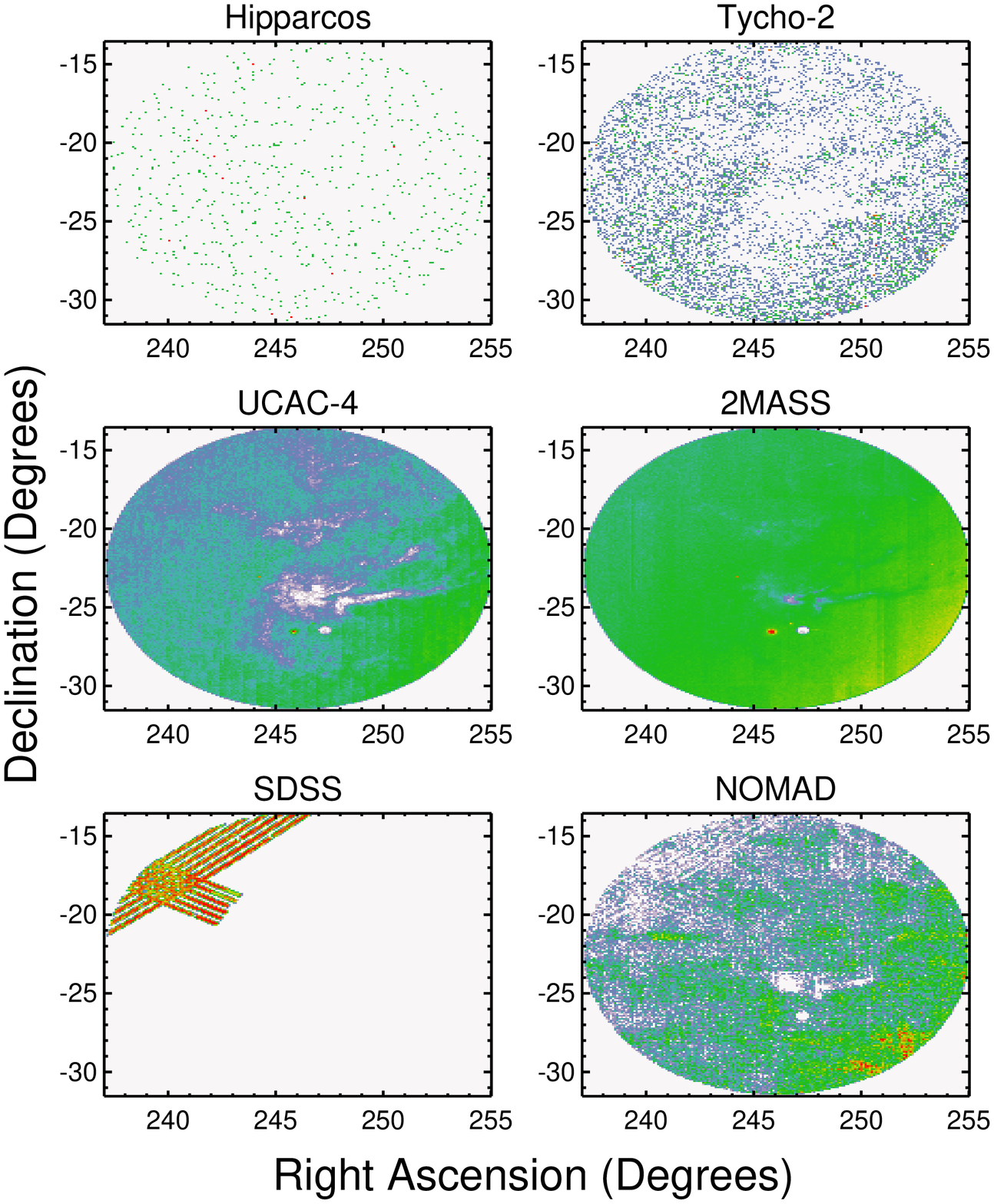}}\hfill
\resizebox{8.75cm}{!}{\includegraphics{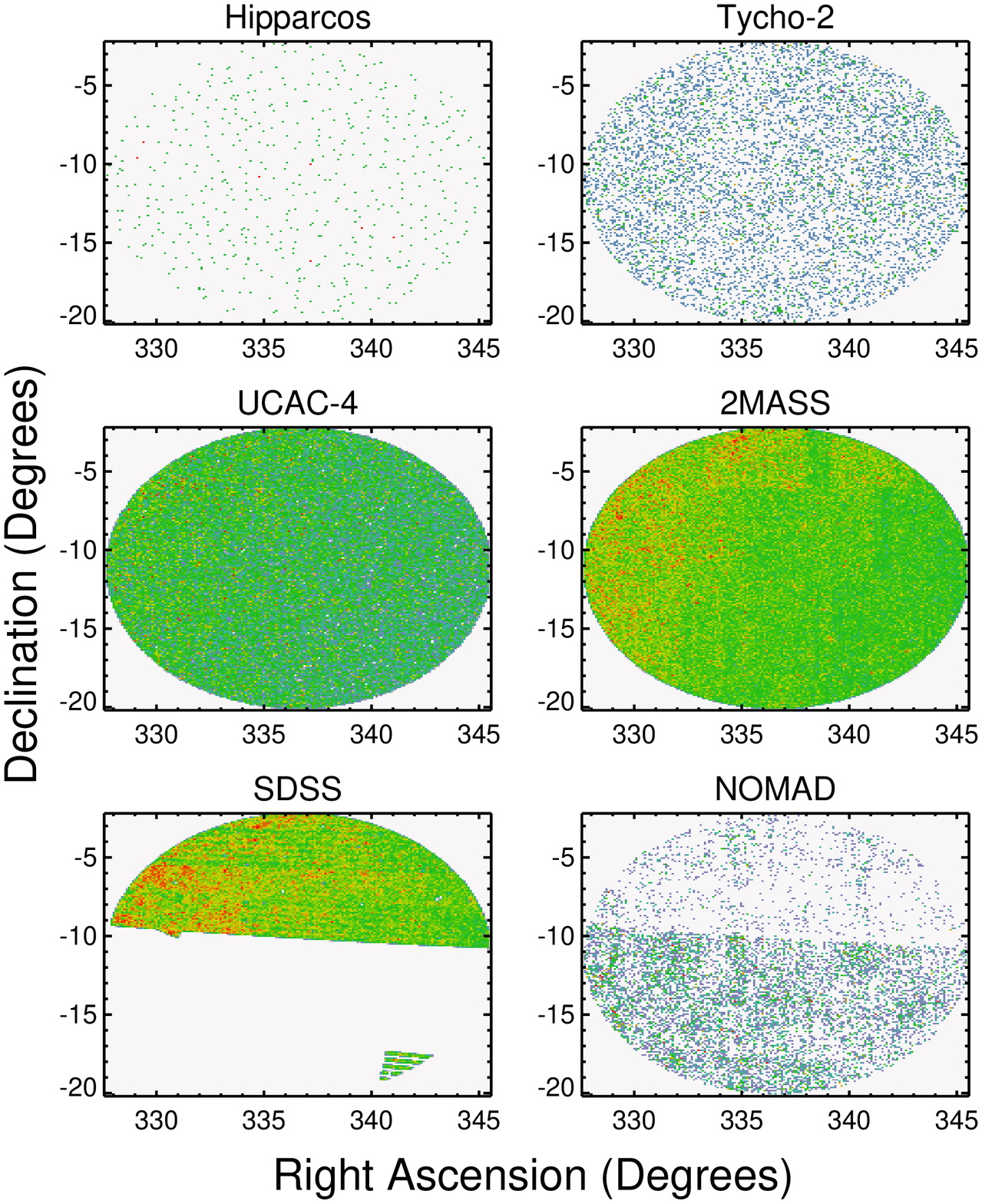}}
\caption{Distribution of EPIC sources in C2 (left) and C3 (right) 
subdivided into the different input catalogs. Color coding denotes logarithmic number 
density in each field. Note that 
the lack of sources near the field center for C2 is due to dust extinction from the 
$\rho$\,Oph star forming region.  The hole at 
RA$\approx 247.4$ degrees and Dec$\approx -26.4$ degrees is caused by $\alpha$ Sco 
($V \approx 0.9$ mag). The bimodal distribution in the 
bottom right panel for C3 is caused by the fact that NOMAD proper motions are only used for 
2MASS cross-matched sources that do not have proper motions listed in Tycho, UCAC or SDSS.}
\label{fig:epiccoverage}
\end{center}
\end{figure*}

For each source, the following observables were cataloged (see Table \ref{tab:epic}): Right 
Ascension (RA), Declination (DEC), proper motion 
in RA, proper motion in DEC, parallax, 
Johnson $BV$, 2MASS $JHK$, Sloan $ugriz$, and associated uncertainties.  For sources 
with photometry in multiple catalogs, Tycho $BV$ and UCAC $gri$ were prioritized over 
APASS $BV$ and SDSS $gri$ to avoid saturation problems for bright stars. 
Cross-matched identifiers of all catalogs were also recorded. 
Additionally, extended sources identified in either 2MASS or SDSS were flagged in 
the ''Objtype'' identifier. 

Following community advocacy, starting with C2 the EPIC was supplemented with proper motions 
from the NOMAD1 catalog \citep{zacharias05} to 
increase completeness for faint high-proper motion stars. NOMAD values were only used for 
2MASS cross-matched sources for which no other proper motions were available (i.e., NOMAD 
does not override Tycho, UCAC, or SDSS). Additionally, 2MASS quality and 
proximity flags were added. We refer the reader to \citet{skrutskie06} 
for the documentation on these flags.

Figure \ref{fig:epiccoverage} shows the sky coverage of each input 
catalog for C2 and C3. Note that SDSS DR9 frequently does not cover the 
entire field, and hence the completeness of the EPIC may vary within a given 
campaign. Additionally, individual catalogs may show local incompleteness 
due to bright stars or regions with high extinction, which are clearly visible in 
Figure \ref{fig:epiccoverage} for the optical catalogs in C2.

\subsection{EPIC ID Assignment}

Each catalog source is assigned a unique EPIC number, which serves as the K2 identifier. 
Table \ref{tab:epicids} lists EPIC IDs assigned for sources from C0--C13. Catalogs 
are divided into 2 types: ``Comprehensive catalogs'' 
are base catalogs for each campaign, and EPIC IDs are assigned with increasing  
declination. ``Missing targets'' are sources that 
were proposed for K2 observations by the community but were not present in the comprehensive 
catalog, for example due to being fainter than the completeness limit. 
For such cases, the K2 Science Office assigned new EPIC IDs and added the sources to the 
catalog for target management. We note that such sources only have coordinates and 
\kep\ magnitudes as provided by the community, but lack ancillary 
information such as magnitudes and cross-matched identifiers that are typically available for other 
EPIC sources.

The EPIC ID bears no information on the campaign 
in which a target has been observed. This is due to the fact that K2 will 
re-observe parts of some fields, and hence a given target can be contained in more than 
one campaign. EPIC IDs start at 201 Million, while K2 custom aperture 
targets have IDs between 200 million and 201 million. Table \ref{tab:ids} provides an 
overview of the ID ranges used for different targets in the \kep\ and K2 
missions. 

\begin{table*}
\begin{small}
\begin{center}
\caption{Log of EPIC IDs assigned by Campaign}
\begin{tabular}{c c c c c r}        
\hline         
Delivery Date &	Campaign & Catalog Type & First EPIC ID & Last EPIC ID & Number of Targets \\
\hline         
02/05/14	&	C1	&	Comprehensive		&	201000001   &	202059065 &    1059065  \\
02/10/14	&	C0	&	Proposed Targets	&	202059066   &	202154323 &    95258	\\
04/07/14	&	C2	&	Comprehensive		&	202154324   &	205871527 &    3717204  \\
04/07/14	&	C3	&	Comprehensive		&	205871528   &	206598205 &    726678	\\
04/18/14	&	C0	&	Comprehensive		&	206598206   &	210282463 &    3684258  \\
04/25/14	&	C1	&	Missing Targets		&	210282464   &	210282491 &    28	\\
07/03/14	&	C2	&	Missing Targets		&	210282492   &	210282560 &    69	\\
07/03/14	&	C4	&	Comprehensive		&	210282561   &	211233315 &    950755	\\
07/03/14	&	C5	&	Comprehensive		&	211233316   &	212235315 &    1002000  \\
09/09/14 	&	C3	&	Missing Targets		&	212235316   &	212235356 &    41		\\
09/14/14 	&	C6	&	Comprehensive		&	212235357   &	212886326 &    650970		\\
09/17/14 	&	C7	&	Comprehensive		&	212886327   &	220115896 &    7229570		\\
03/30/15 	&	C4	&	Missing Targets		&	220115897   &	220115973 &    77		\\
03/30/15 	&	C8	&	Comprehensive		&	220115974   &	220769262 &    653289		\\
03/30/15 	&	C9	&	Comprehensive		&	220769263   &	228682308 &		7913046		\\
03/30/15 	&	C5	&	Missing Targets		&	228682309   &	228683400 &		1092		\\
03/30/15 	&	C10	&	Comprehensive		&	228683401   &	229227137 &		543737		\\
12/03/15        &       C6      &       Missing Targets		&	229227138   &	229228144 &		1007		\\
12/03/15        &       C7      &       Missing Targets		&	229228145   &	229228373 &		229		\\
12/03/15        &       C8      &       Missing Targets		&	229228374   &	229228998 &		625		\\
12/03/15        &       C11     &       Comprehensive		&	229228999   &	245899083 &		16670085	\\
12/03/15        &       C12     &       Comprehensive		&	245899084   &	246541209 &		642126	        \\
12/03/15        &       C13     &       Comprehensive		&	246541210   &	248368655 &		1827446	        \\
\hline
\end{tabular} 
\flushleft Notes: ``Comprehensive'' are base catalogs including all 
information listed in Table \ref{tab:epic}.  ``Proposed Targets'' and ``Missing Targets'' are 
sources generated based on community-provided information and hence do not 
include catalog cross-matched information such as alternate identifiers and photometry 
(see Section \ref{sec:c0adj} for details). Note that the completeness of the EPIC in C9 is reduced 
due to the high source density (see Section \ref{sec:c9}).
\label{tab:epicids}
\end{center}
\end{small}
\end{table*}

\begin{table}
\begin{small}
\begin{center}
\caption{Identifiers used by the Kepler and K2 Missions}
\begin{tabular}{l l}
\hline
ID &	Targets \\
\hline
0-30M		&		KIC (\kep\ Catalog Targets)		\\
30M-60M		&		UKIRT KIC extension targets			\\
60M-100M	&		K2 engineering test targets			\\
100M-200M	&		\kep\ Custom Aperture Targets		\\
200M-201M	&		K2 Custom Aperture Targets			\\
$>$201M		&		EPIC (K2 Catalog Targets)    		\\
\hline												     
\end{tabular} 
\label{tab:ids}
\end{center}
\end{small}
\end{table}

\section{Kepler Magnitudes}
\label{sec:kpmags}

\kep\ magnitudes (\textit{Kp}) are estimated using observed broadband magnitudes. To 
place K2 \textit{Kp} values as closely as possible to \kep\ \textit{Kp} values, 
we follow the definitions using 
$gri$ photometry in Equations (2)--(5) by \citet{brown11}. However, 
since $gri$ photometry is not available for every source, different methods had to be adopted.
The EPIC identifier ``KepFlag'' keeps track of which photometry 
was used to calculate the \kep\ magnitude.  The following prioritization 
was adopted (brackets denote the Kepflag string listed in Table \ref{tab:epic}):

\begin{list}{\labelitemi}{\leftmargin=1em}

\item Kepflag = [gri]: \textit{Kp} was calculated from $gri$ magnitudes using Equations 
(2)--(5) in \citet{brown11}.

\item Kepflag = [BV]: Sloan $gr$ was estimated from Johnson $BV$ using the transformations by 
\citet{bilir05}:

\begin{equation}
g-r = 1.124 (B-V) - 0.252 \: ,
\end{equation}

\noindent
and 							

\begin{equation}
g = V + 0.634 (B-V) - 0.108 \: .	
\end{equation}

\textit{Kp} was then calculated from $gr$ using Equation (2) in \citet{brown11}.

\item Kepflag = [JHK]:  \textit{Kp} was calculated using the polynomial $J-K$ relations by 
\citet{howell12}.  Given $x=J-K$ these transformations are:

\begin{equation}
\begin{split}
Kp = 0.42443603 + 3.7937617 x - 2.3267277 x^{2} \\ + 1.4602553 x^{3} + K \: ,
\end{split}			
\end{equation}

for all stars with $J-H > 0.75$ and $H-K > 0.1$ (approximate color cut for giants), and 

\begin{equation}
\begin{split}
Kp = 0.314377 + 3.85667 x + 3.176111 x^{2} - \\ 25.3126 x^{3} + 40.7221 x^{4}  \\ 
- 19.2112 x^{5} + K \: ,	
\end{split}			
\end{equation}

for all remaining stars.  The above relations are applied for $-0.2 < J-K < 1.2$ for 
giants and $-0.2 < J-K < 1.0$ for dwarfs, which are the calibration ranges given by 
\citet{howell12}.

\item Kepflag = [J]: For sources outside the color limits of Equations (3) and (4) or 
sources 
which only have a valid $J$-band magnitude, a rough estimate of \textit{Kp} was calculated from the 
$J$-band magnitude using the relations by \citet{howell12}:

\begin{equation}
\begin{split}
Kp = -398.04666 + 149.08127 J - 21.952130 J^{2} + \\ 1.5968619 J^{3} - 0.057478947 J^{4}  \\ 
+ 0.00082033223 J^{5} + J \: ,	
\end{split}			
\end{equation}

for $J = 10 -16.7$ and 

\begin{equation}
Kp = 0.1918 + 1.08156\,J 
\end{equation}

for $J > 16.7$. 

For bright near-infrared sources ($J<10$) which do not have optical photometry 
and fall outside the $J(HK)$ calibration range in \citet{howell12}, \kep\ magnitudes 
were assumed to be equal to their $J$-band magnitude for C1--C7. Starting with 
C8, this criterium was changed to $Kp=J+1.7$ to avoid a 
discontinuity in \kep\ magnitudes for highly reddened fields in which 
a significant fraction of sources lack optical photometry and fall outside the 
calibration range \citep[see Figure 18 in][]{howell12}. We note that this discontinuity 
only affects $<0.1\%$ of all sources in the EPIC for C1--C7.

\end{list}

\begin{figure}
\begin{center}
\resizebox{\hsize}{!}{\includegraphics{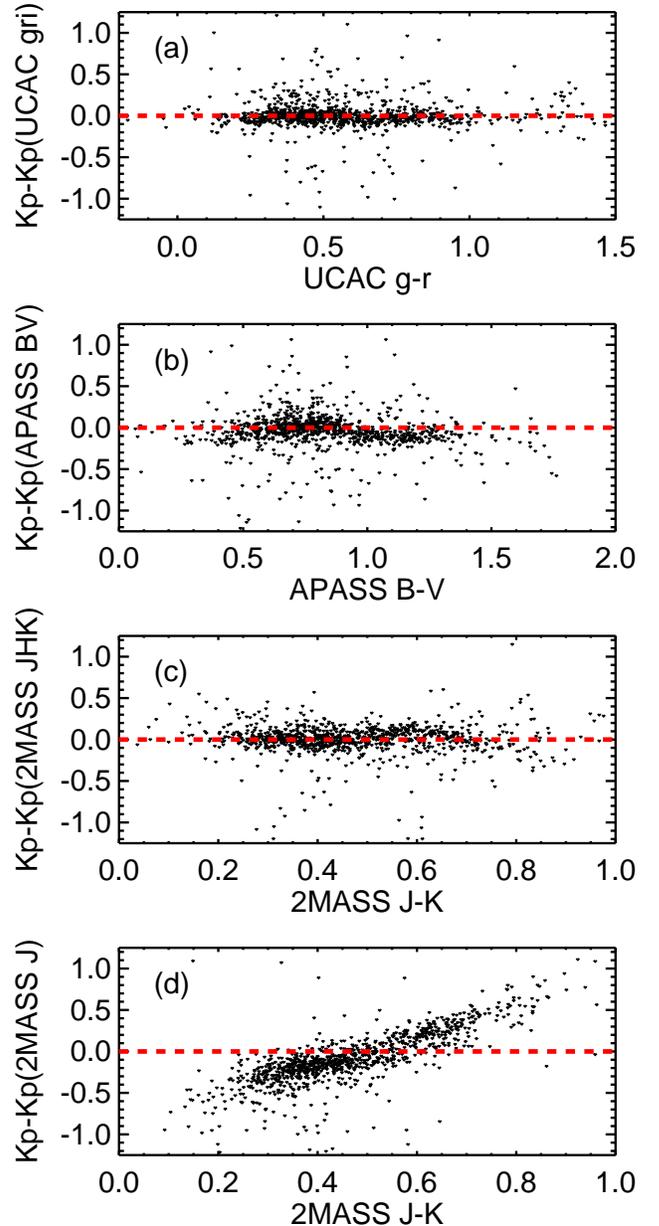}}
\caption{Difference between \kep\ magnitudes calculated from UCAC $gri$ (panel a), 
APASS $BV$ (panel b), 2MASS $JHK$ (panel c), and 2MASS $J$ (panel d) as a function of 
color for 5000 random targets in the original \kep\ field.}
\label{fig:testkepmag}
\end{center}
\end{figure}

\noindent
Figure \ref{fig:testkepmag} compares \kep\ magnitudes calculated from UCAC $gri$, APASS $BV$, 2MASS $JHK$, and 
2MASS $J$ to original \textit{Kp} values for a random sample of 5000 targets in the original \kep\ field.  
The values show good agreement for the first three methods (panels a--c), with a median 
offset and scatter of $-0.01\pm0.09$ mag for UCAC $gri$, 
$-0.03\pm0.13$ mag for APASS $BV$, and $0.01\pm0.12$ mag for 2MASS $JHK$.  Users should 
be aware that 
\kep\ magnitudes based on $J$ (panel d) are very approximate since the transformation is 
based on the average colors of stars in the \kep\ field.  Deviations of up to 1 mag can 
be observed for very blue or red sources.

\begin{figure}[h!]
\begin{center}
\resizebox{\hsize}{!}{\includegraphics{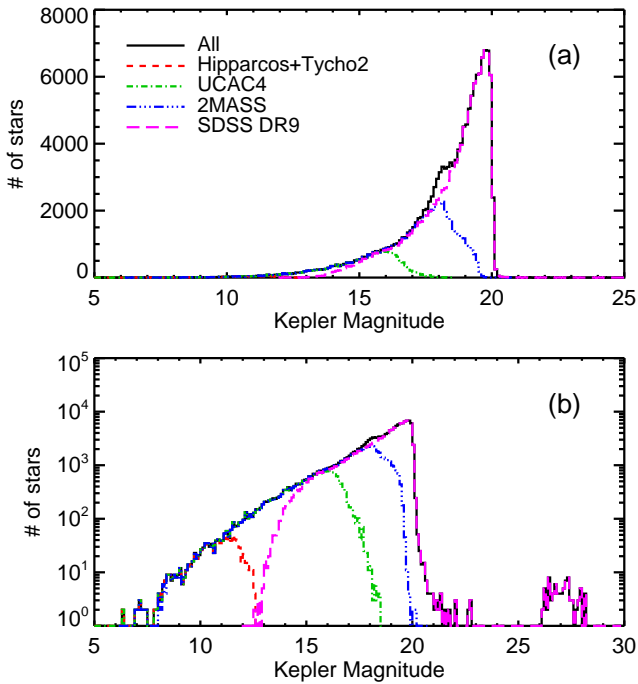}}
\caption{Histogram of EPIC \kep\ magnitudes for sources in a $\approx$\,80 square degree field in 
C1 covered by both SDSS and 2MASS on a linear scale (panel a) and logarithmic 
scale (panel b).  Colors show the individual contributions from different catalogs. 
The completness limit of the EPIC is $Kp \approx 19$ for areas covered by SDSS, 
and $Kp \approx 17$ for areas covered only by 2MASS. }
\label{fig:kphisto}
\end{center}
\end{figure}

Figure \ref{fig:kphisto} shows the \kep\ magnitude distribution for sources in a $\approx$\,80 
square degree field 
in C1 covered by 2MASS and SDSS.  The steep drop-off at Kp $\approx$ 20 mag is 
caused by the $r < 20$ mag cut in SDSS DR9.  For regions not covered by SDSS the 
completeness is set by 2MASS. Panel (b) of Figure \ref{fig:kphisto} shows the 
distribution on a logarithmic scale, illustrating that the catalog includes a small 
number of objects with $Kp > 25$, which are predominantly galaxies identified in SDSS.

Figure \ref{fig:kphisto} and visual inspection of POSS-II images showed that the EPIC is 
complete to about 
$Kp \approx 19$ for areas covered by SDSS, and $Kp \approx 17$ for areas covered 
by 2MASS only. Completeness for fields only covered by 2MASS may be significantly 
reduced for sources that are blue and faint.

\section{Catalog Adjustments and Known Shortcomings} 

\subsection{Merging and Overrides for Missing Targets}
\label{sec:c0adj}

Targets which were proposed for observations by the community but not 
matched with an existing EPIC ID are reconciled with the comprehensive catalog 
for each campaign to avoid source duplication and consequent complications for 
computing source crowding, flux fraction, and optimal apertures \citep[see][]{archivemanual}. 
The following section describes this procedure for C0, 
an engineering run for which the this reconciliation 
was particularly important since
target selection began before an EPIC was generated. The procedures described 
below are also used for later campaigns, and hence apply to most 
proposers who submitted targets without an existing EPIC ID.

\begin{figure*}
\begin{center}
\resizebox{6.01cm}{!}{\fbox{\includegraphics{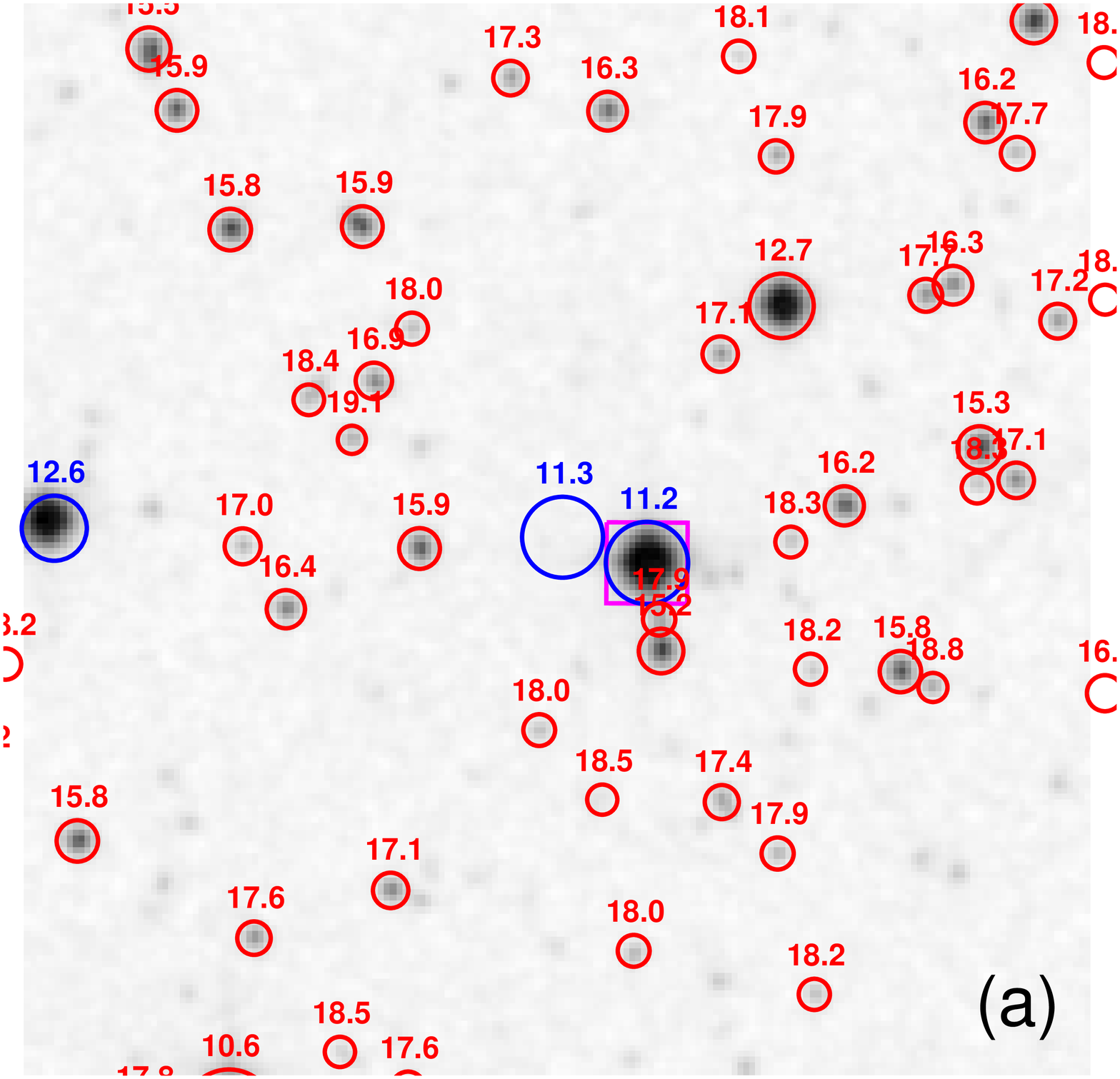}}}\hfill
\resizebox{5.85cm}{!}{\fbox{\includegraphics{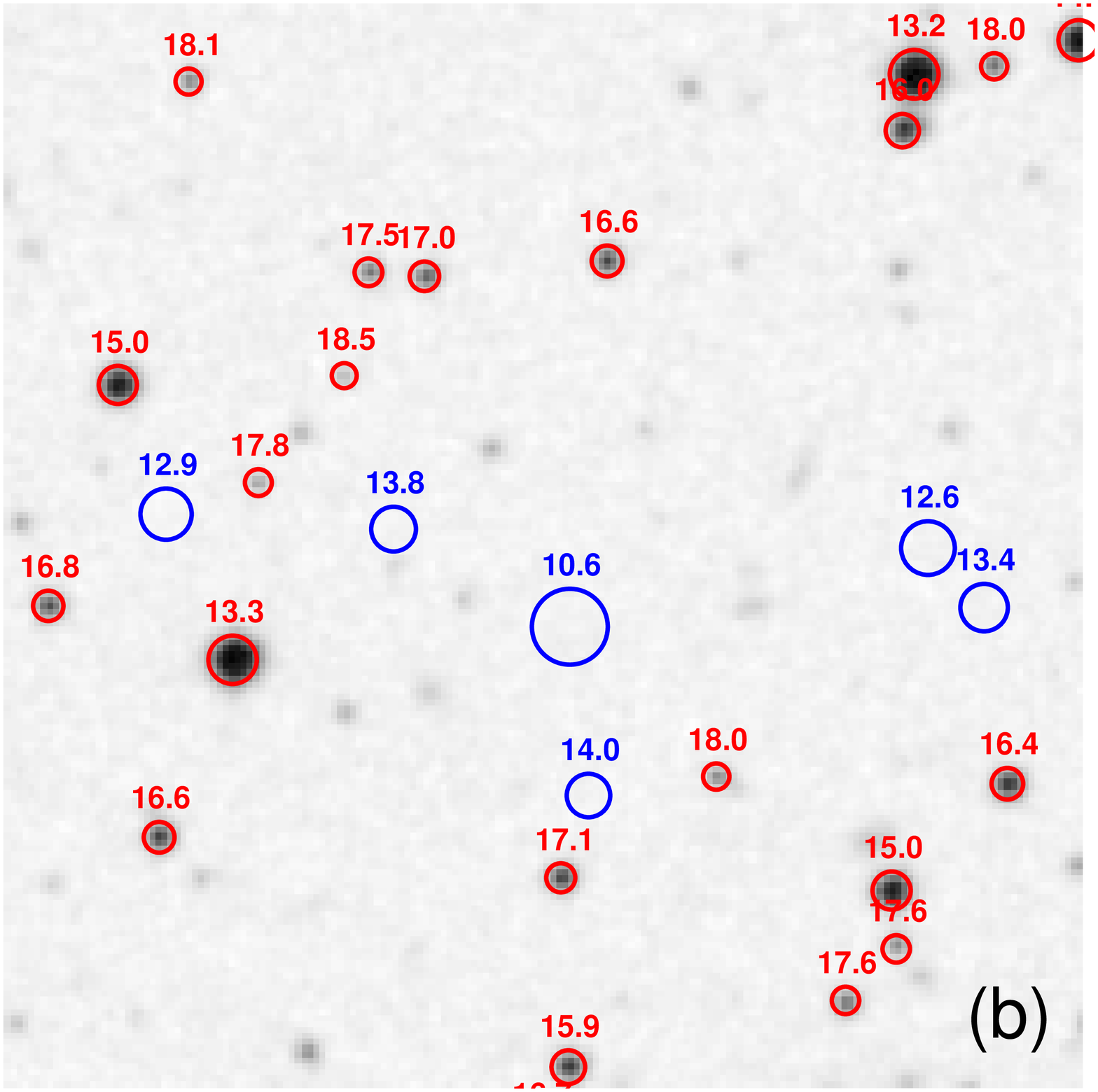}}}\hfill
\resizebox{5.85cm}{!}{\fbox{\includegraphics{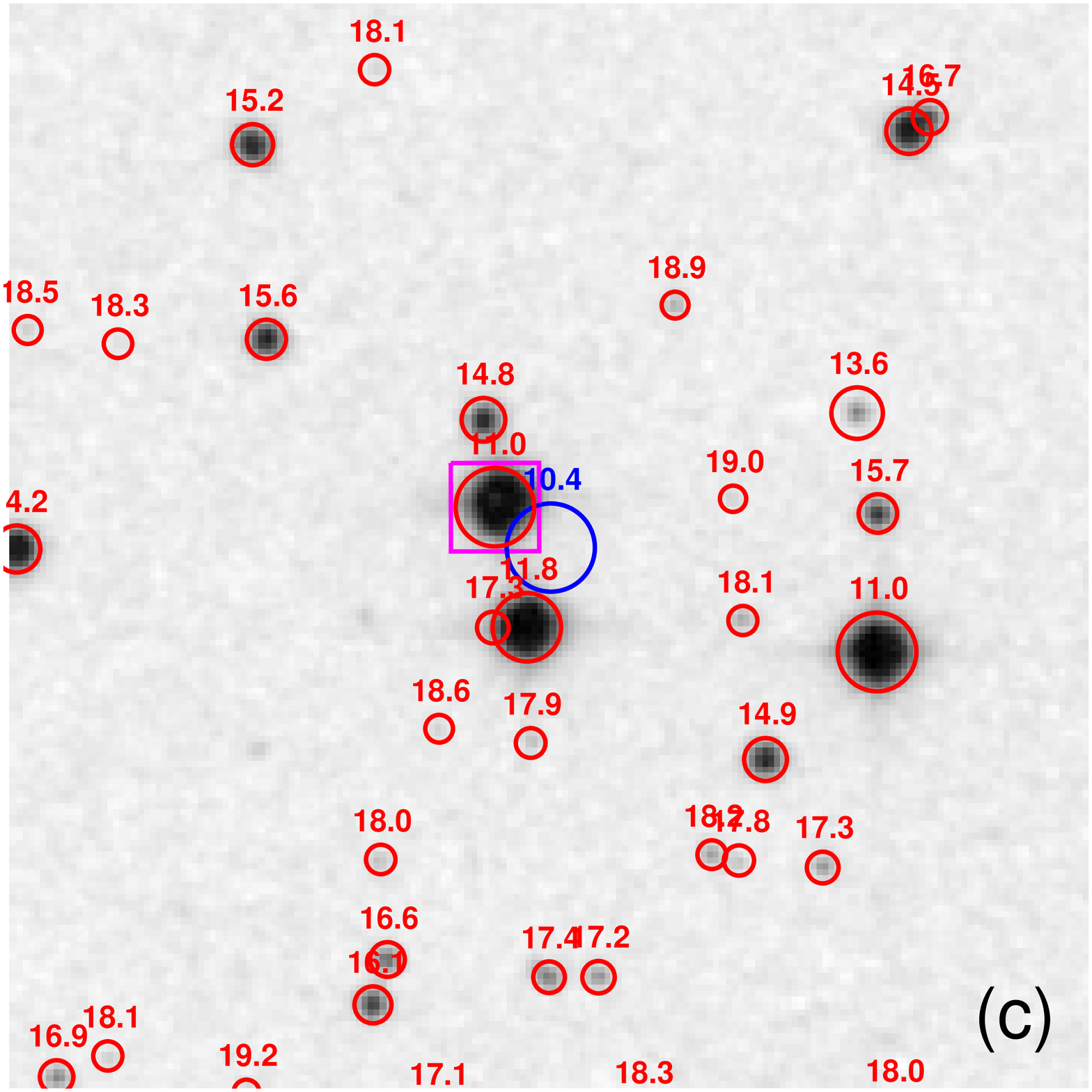}}}
\caption{POSS-2 images in the C0 field with EPIC sources from the Comprehensive Catalog 
overlaid (red circles). Each image spans $3\times3$ arcminutes and circle 
sizes scale with 
\kep\ magnitude, which is given next to each source. Each panel demonstrates 
cases of reconciling sources provided by the community (Target Catalog, 
blue circles) with the EPIC: 
duplicate sources in the Target Catalog (panel a),
erroneous sources in the Target 
Catalog (panel b),
and source merging between the Target 
and Comprehensive Catalog (panel c).
Magenta squares in panels a and c show the adopted source.}
\label{fig:dss}
\end{center}
\end{figure*}

For C0, EPIC IDs were assigned to the 7748 targets approved for flight during the 
proposal review and target management process. This set of observed targets is 
referred to below as the ``Target Catalog''. The only columns populated in this 
catalog were those provided by the proposers: \kep\ magnitude, right ascension, and 
declination. Subsequently, a ``Comprehensive Catalog'' was created
and merged with the Target Catalog. Sources were considered matched if their positions 
agreed to within $<$\,10 arcseconds and their \kep\ magnitudes agreed to within 
$<$\,4 magnitudes.  For 
matched sources, the Target Catalog was given precedence over the Comprehensive 
Catalog, therefore trusting the positions and magnitudes provided by successful 
proposers. This means that all C0 targets do not contain 
cross-matched identifiers or broadband magnitudes as described in 
Section \ref{sec:xmatching} in the EPIC\footnote{Following community advocacy, 
the C0 EPIC cross-matches have been made available as a separate delivery at 
MAST (\url{http://archive.stsci.edu/k2/catalogs.html}). We caution 
that it cannot be guaranteed that the cross-matched information corresponds to the 
intended targets proposed by the community.}, and that 
the listed \kep\ magnitudes were calculated differently than described in 
Section \ref{sec:kpmags}.

A total of 137 sources in the C0 Target Catalog had no match in the Comprehensive Catalog 
for C0, and underwent closer inspection. For these unmatched sources, the following 
procedure was adopted:

\begin{itemize}
\item[(1)] If $Kp>14$ (80 targets), unmatched Target Catalog sources were 
included with no 
change because the Comprehensive Catalog is significantly incomplete at faint magnitudes. 

\item[(2)] If $Kp<14$ (11 targets) and the Target Catalog had a duplicate source, then the 
source that best agreed with the Comprehensive Catalog was included and the nearby 
duplicate was forced to  $Kp=30$, thereby declaring it an artifact.  
An example of this situation is shown in Figure \ref{fig:dss}a. In such cases, the K2 
photometry pipeline is 
expected to produce valid photometry for the first source, but will only export 
calibrated pixel-level data for the second. Consequently, proposers who submitted poor 
source coordinates might find more completely processed data under an alternate EPIC ID.

\item[(3)] If $Kp<14$, the Target Catalog has no duplicate source, and:

\begin{itemize}
\item[(a)] $Kp>12$ (23 targets), then the magnitude of the source was forced to $Kp=30$. 
Several examples are shown in Figure \ref{fig:dss}b. The rationale is that the Comprehensive Catalog 
is expected to be complete at these magnitudes, so unmatched sources are likely
artifacts. 

\item[(b)] $Kp<12$ (2 targets), and a corresponding source was found in the Comprehensive 
Catalog within a distance $< 30''$, then the Target Catalog source was replaced by its 
match in the Comprehensive Catalog.  An example is shown in Figure \ref{fig:dss}c. 
The rationale 
is that for these bright sources uniqueness is less of an issue, so the search space 
was expanded to obtain matches that led to more reliable parameters.  For C0 the 
aperture masks included a large halo of 10 pixels (40'') in all directions, 
and hence the requisite pixels for the intended target were likely collected. 

\item[(c)] $Kp<12$ (14 targets), and no corresponding source was found in the 
Comprehensive 
Catalog within $< 30''$, then the magnitude of the source was forced to 
Kepmag = 30.  The rationale is that larger offsets probably push the proposed targets 
outside their aperture masks even if the intended sources could be correctly identified 
in the Comprehensive Catalog.
\end{itemize}

\item[(4)] The remaining (7) targets were included because they were judged to be plausible 
omissions or magnitude mismatches in the Comprehensive Catalog 
based on visual checks with POSS-II images.
\end{itemize}

Details on C0 Target Catalog overrides can be found in the 
online EPIC documentation\footnote{\url{http://archive.stsci.edu/k2/epic.pdf}}. 
Similar corrections as described in this section 
were also applied to proposed targets without a cross-matched 
EPIC ID after C0.

\subsection{Duplicate Sources for Bright Stars with Infrared Excess and 
Faint SDSS Stars}

Matches between Hipparcos and 2MASS are based on position and apparent 
magnitudes, the latter being estimated through empirical 
conversions from 2MASS bands to $V$. For stars with strong infrared excess 
(such as young stars with debris disks), the conversion from $JHK$ to $V$ can 
produce wrong $V$-band estimates, causing the EPIC code to erroneously 
generate two bright sources since the apparent magnitude match fails. 
An example is T\,Tauri, which was catalogued with 
two EPIC IDs (EPIC\,210777988 and EPIC\,210777987).
In general, such errors should be rare since such optical-infrared 
magnitude matches are restricted to Hipparcos and Tycho.

Similar to the Hipparcos-2MASS cross-match, sources between 2MASS and SDSS are 
matched using positions and apparent magnitudes by applying transformations from 
$JHK$ to Sloan $g$. If no 2MASS $HK$ photometry with quality flag better than C is 
available, this magnitude match is dropped and only a position match is required. 
Due to a coding error this criterion was not correctly 
applied for C0--C7, and hence a fraction of faint sources may have erroneously 
duplicated EPIC sources generated from the same 2MASS and SDSS source.

\subsection{Missing NOMAD Proper Motions}
Following community advocacy, starting from C2 proper motions in the EPIC were 
supplemented by values provided in the NOMAD1 catalog \citep{zacharias05}. These 
overrides were intended 
to be only applied for sources that do not have proper motions from the other 
input catalogs (i.e. Tycho, UCAC, and SDSS). Due to a coding error this 
criterion was not correctly applied in C3--C7, and consequently for some Tycho/UCAC/SDSS stars 
proper motions were replaced with NOMAD values, or were left blank although 
NOMAD proper motions exist. Identifiers were in all cases propagated correctly, 
and hence all sources with a NOMAD identifier have their proper motion values 
taken from the NOMAD catalog.

\subsection{Missing Sources due to Quality Flag Cuts}
In rare cases the quality cuts applied for 2MASS photometry may remove genuine 
infrared bright sources with no optical counterparts. One example for this case is 
QQ\,Tau, which has no $BVgri$ photometry in UCAC4 and a $J$-band E quality flag. 
Since 2MASS $HK$ are not sufficient to calculate a \kep\ magnitude the 
source was dropped from the EPIC, despite its bright $J$-band magnitude. QQ\,Tau was 
consequently added to the EPIC after being successfully proposed for observations in 
C4, following the procedures outlined in Section \ref{sec:c0adj}.

\subsection{Regions of Incompleteness}
For C7 there was no 2MASS data available for a rectangular 
region about 0.1 degrees in right ascension by 0.3 degrees in declination 
that is centered near RA = 287.9$^{\circ}$, 
Dec = -17.7$^{\circ}$ and for a circular region about 0.07$^{\circ}$ in diameter 
that is centered near RA = 285.3$^{\circ}$, Dec = -22.7$^{\circ}$. While the 
latter is almost certainly due to masking associated with the 8.2 magnitude star at the 
center of the region, the former is not understood and sky images do show 
bright sources in that area. 
The result is that the EPIC is much 
less complete in these two regions than in the remainder of the C7 field of view. These 
regions are relatively small compared to the region caused by $\alpha$\,Sco 
in the C2 (see Figure \ref{fig:epiccoverage}). 
We emphasize that regions of incompleteness likely exist for 
most campaigns, and hence users should be aware that certain photometric bands in the 
EPIC may be affected by this.

\subsection{Reduced Completeness for C9}
\label{sec:c9}
Because C9 is near the galactic plane and includes portions of the galactic bulge, 
the stellar density is at least a factor of ten higher than for other campaigns.  
Consequently, the 2MASS input catalog was truncated at $J<14$ in order to limit the number of 
catalog entries for C9 to 7.9 million for computational reasons (see 
Table \ref{tab:epicids}).  Because UCAC4 is typically complete to $Kp\approx15$, 
the EPIC is likely complete to about this magnitude, but the resulting estimates are less 
accurate due to large reddening in this field. 
Additionally, due to differential reddening between optical and infrared 
magnitudes, \textit{Kp} values derived from $JHK$ will be underestimated (too bright)
compared to \textit{Kp} values derived from optical bands. 
Fortunately, these issues are expected to have little impact on the aperture sizes used 
for data collection, particularly since most pixels collected during C9 
will reside in large super apertures to facilitate a microlensing 
experiment \citep{henderson15}. Due to code improvements 
comprehensive catalogs in crowded fields following C9 are no longer magnitude limited, 
as can be seen by the C11 EPIC which contains $\approx$\,17 million sources.

\subsection{Systematic Offsets in \kep\ Magnitudes}
Early K2 campaigns revealed that EPIC \textit{Kp} values derived from 
2MASS photometry (Kepflag='JHK' or Kepflag='J', see Section \ref{sec:kpmags}) 
are systematically too low (too bright) by about one magnitude compared to flux 
measurements from the K2 data\footnote{\url{http://keplerscience.arc.nasa.gov/K2/C4drn.shtm}}. 
Further analysis showed that the majority of the affected targets are faint ($Kp>14$) and 
red ($J-K>0.8$), consistent with M dwarfs. We were unable to reproduce this offset 
by comparing \textit{Kp} calculated from $JHK$ with \textit{Kp} values in the KIC, 
or by comparing \textit{Kp} calculated from $JHK$ with 
\textit{Kp} values calculated from APASS $gri$ in C4.
This indicates that the offset is likely due to a discrepancy in the definition 
of \textit{Kp} compared to actual flux measurements by the spacecraft for red stars, 
which has so far went unnoticed due to relatively small number of red stars 
targeted in the \kep\ field. However, 
because the offset causes the K2 apertures to be too large, this should not 
significantly affect science investigations for these targets.

A second population of stars in C4 showed \textit{Kp} values which are systematically too high 
(too faint) compared to actual flux measurements. In contrast to targets with underestimated 
magnitudes, this sample appeared localized in certain regions of the K2 field of view. 
Further analysis showed that the erroneous \textit{Kp} values are caused by 
$i$-band values with large uncertainties which are systematically too high, 
originally adopted from APASS. Future catalogs will mitigate this problem by disregarding photometry 
with large uncertainties when calculating \kep\ magnitudes.

\section{Stellar Characterization}

Unlike the KIC for the \kep\ Mission, the EPIC does not include 
stellar properties for each source.
Consequently, target selection for K2 has relied mostly on color and 
proper motion cuts. The most comprehensive source for stellar classifications 
of K2 targets to date is the TESS target selection catalog \citep{stassun14}.  
As of yet, however, this catalog only includes luminosity classes and effective 
temperatures, and does not take into account reddening which can lead to 
biases in the effective temperature scale.
In this section, we derive a full set of stellar properties (temperatures, 
surface gravities, metallicities, radii, masses, densities, distances, and extinction) for 
K2 targets to investigate the target populations, and provide the community with a 
resource for a first-look characterization of K2 targets.
We do not classify all EPIC sources due to computational 
reasons and to keep the community target selection process free of biases which are 
inherently present for model-based stellar parameter inference.

\subsection{Apparent Magnitude Distribution of K2 Targets}

\begin{figure}
\begin{center}
\resizebox{\hsize}{!}{\includegraphics{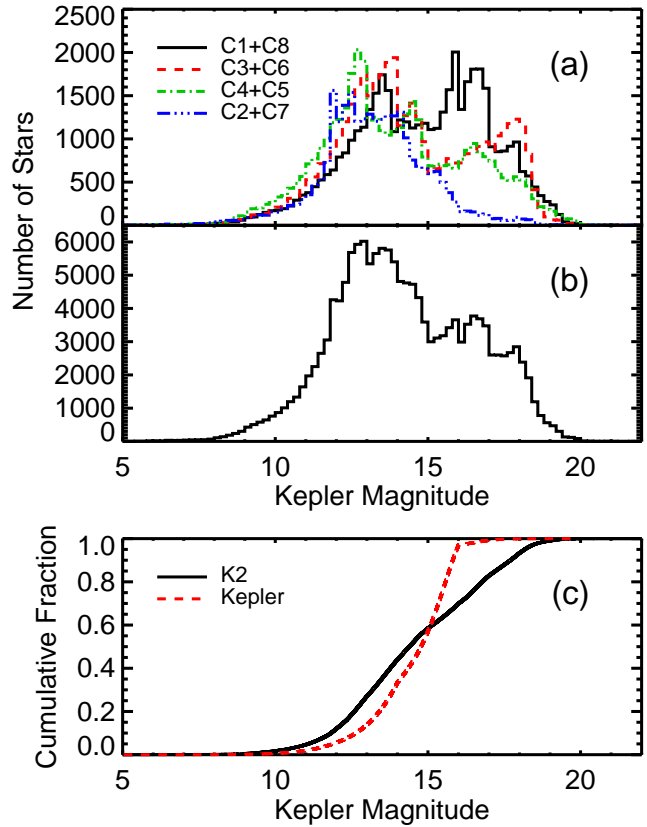}}
\caption{
Panel (a): EPIC \kep\ magnitude distribution for \ntotkp\ 
K2 targets in C1--C8, split into campaigns with similar heights above the 
galactic plane: 
$|b|\approx 57$--$58^{\circ}$ (C1+C8),
$|b|\approx 50$--$52^{\circ}$ (C3+C6),
$|b|\approx 26$--$32^{\circ}$ (C4+C5),
$|b|\approx 15$--$19^{\circ}$ (C2+C7).
Panel (b): \kep\ magnitude distribution of K2 targets over all campaigns. Panel (c): 
Cumulative \kep\ magnitude distribution of K2 targets (black solid line) compared to the 
\kep\ target sample (red dashed line).}
\label{fig:maghisto}
\end{center}
\end{figure}

\begin{figure*}
\begin{center}
\resizebox{\hsize}{!}{\includegraphics{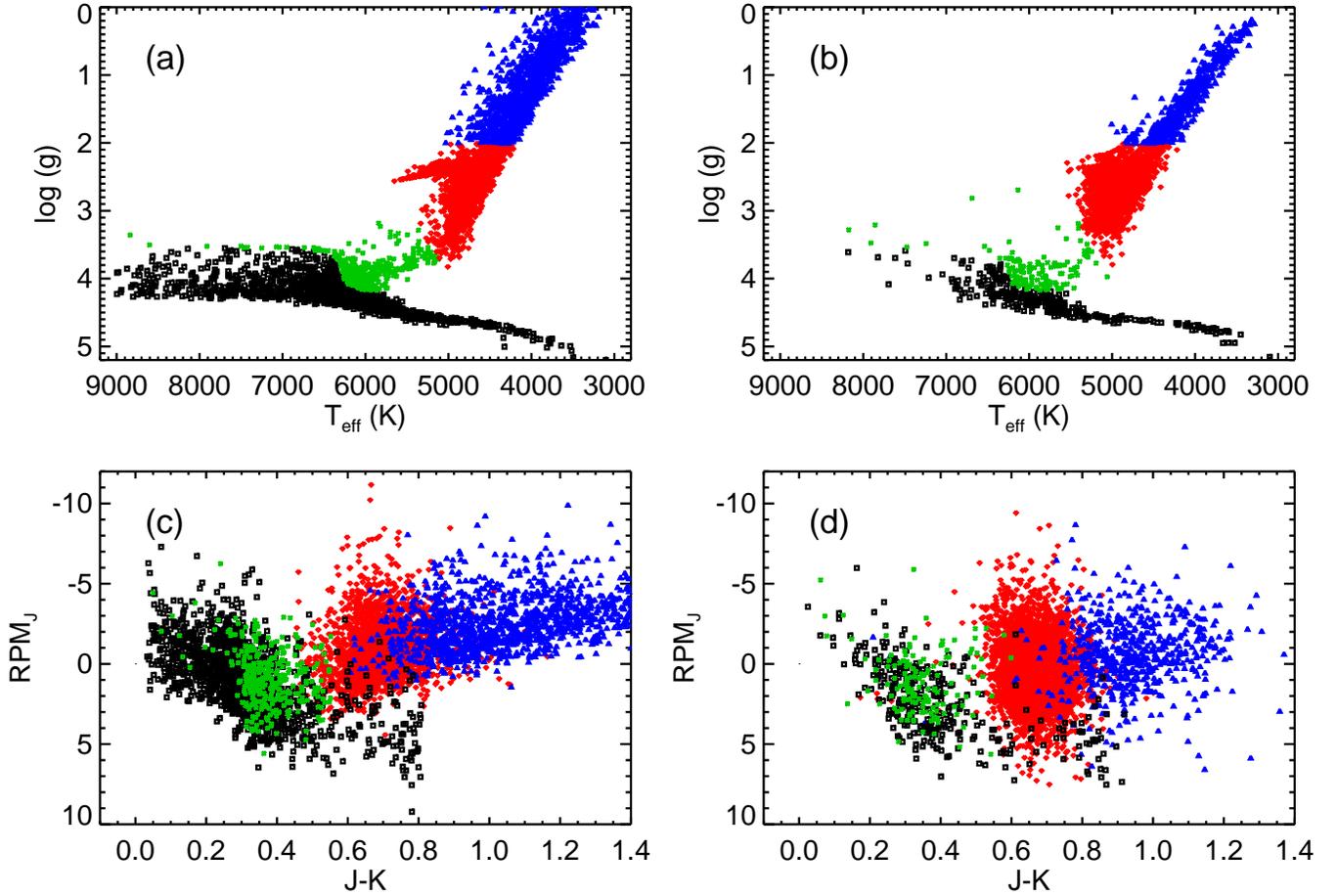}}
\caption{\teff\ versus \logg\ for a synthetic population 
in the \kep\ field generated using \textit{Galaxia} (panel a) with an apparent magnitude 
distribution which matches a control sample of \nkepstars\ \kep\ targets with well determined 
fundamental properties from asteroseismology and spectroscopy (panel b). Different 
symbols and colors denote evolutionary states: dwarfs (black squares), subgiants 
(green asterisks), low-luminosity RGB stars (red diamonds) and high-luminosity 
RGB stars (blue triangles).
Panels (c) and (d) show the same samples as in the top panels but now in a 
$J$-band reduced proper motion versus $J-K$ color diagram. Only 5000 stars of the 
full sample are shown for clarity.}
\label{fig:rpmcolor}
\end{center}
\end{figure*}

Figure \ref{fig:maghisto} shows the EPIC \kep\
magnitude distribution for \ntotkp\ K2 targets selected for observations 
in C1--C8, grouped 
into campaigns with similar heights above the galactic plane.
Most campaigns have a 
bimodal distribution as a result of programs targeting
bright FGK dwarfs and red giants for transiting planet searches and galactic 
archeology \citep[e.g.][]{vanderburg15,stello15}, while M dwarf exoplanet 
surveys \citep[e.g.][]{crossfield15} typically 
propose fainter targets. The distributions show that 
fields which are high above the galactic plane (C1, C3, C6, and C8) 
tend to include fainter targets than those close to the plane (C2, C4, C5 and C7). 
Overall, the median apparent magnitude for K2 targets is $Kp\approx14.0$ and $J\approx12.2$. 

Figure \ref{fig:maghisto}c compares the \kep\ magnitudes of K2 targets 
to the distribution of targets in the \kep\ field. 
Interestingly, the median apparent magnitude of K2 targets is 
only $\approx 0.6$\,mag brighter than for 
\kep\ ($Kp\approx14.6$). However, the K2 target list covers a much 
broader apparent magnitude distribution, with a larger number 
of very bright and very faint ($Kp>16$) targets than \kep.

\subsection{Reduced Proper Motions}

For typical apparent magnitudes of K2 targets the only currently available 
data are broadband colors and proper motions. Johnson, Sloan or 2MASS filters 
are sensitive to effective temperatures (\teff), but are 
notoriously insensitive to surface gravities (\logg). For the KIC this problem was 
alleviated through narrow-band $D51$ photometry, which 
provided sensitivity to discern dwarfs from giants. However, $D51$ photometry or 
bluer broadband colors with sensitivity to \logg\ are not 
available in the all-sky surveys used to construct the EPIC.

An alternative for determining evolutionary states are proper 
motions. Taking the $J$-band as an example, reduced proper motions are 
defined as \citep[e.g.][]{gould03}:

\begin{equation}
{\rm{RPM_{J}}} \equiv m_{J}+5\log(\mu) = 
M_{J} + 5\log\left(\frac{v_{t}}{47.4\,\rm{km\,s^{-1}}}\right) \: .
\label{equ:propmotion}
\end{equation}


\noindent
Here, $\rm{RPM_{J}}$ is the reduced proper motion in the J-band, 
$M_{J}$ and $m_{J}$ are the absolute and apparent $J$-band magnitudes, 
$\mu$ is total proper motion in 
arcseconds per year and $v_{t}$ is the transverse velocity in $\rm{km\,s^{-1}}$.
Assuming equal transverse motion of all stars, reduced proper motions are equivalent to 
absolute magnitudes since the apparent motion depends solely on 
the distance to the observer. In reality, stars show significant dispersion in 
$v_{t}$, and hence reduced proper motions are 
only a proxy for absolute magnitudes. Nevertheless, assuming relatively local 
thin disc populations, proper motions can be effective in characterizating stellar 
populations.


To test the usefulness of proper motions 
we first constructed an EPIC for the \kep\ field following the 
procedure described in Section \ref{sec:xmatching}.
We then used \textit{Galaxia} \citep{sharma11} to simulate a 
synthetic stellar population in a $\approx$\,250 square degree circle centered on the 
\kep\ field. \textit{Galaxia} is 
based on the {\sl Besan\c{c}on} model of the galaxy \citep{robin03}, with some modifications. 
It provides model photometry (such as UBV, Sloan and 2MASS),
space velocities (U, V, W), and 3D extinction values 
using a double-exponential disc model with parameters 
optimized to match the dust maps by \citet{schlegel98}, 
along with a correction for high extinction regions as described in \citet{sharma14}.
The isochrones used in \textit{Galaxia} to predict the stellar
properties are from the Padova database \citep{marigo08}.
For each simulated \textit{Galaxia} star we 
calculated proper motions by inverting the equations given by \citet{johnson87}.

Figure \ref{fig:rpmcolor}a shows a $\teff-\logg$ diagram  
for a subsample of the \textit{Galaxia} population in the \kep\ field 
(hereafter referred to as the synthetic sample). The subset was randomly drawn to 
match the apparent $J$-band distribution of a sample of 
\nkepstars\ \kep\ targets with well determined stellar properties from 
asteroseismology and spectroscopy taken from \citet{huber14} (Figure \ref{fig:rpmcolor}b, 
hereafter referred to as the control sample).
Note that the lack of hot stars in Figure \ref{fig:rpmcolor}b compared to 
Figure \ref{fig:rpmcolor}a is due to the \kep\ target selection function. 
To distinguish evolutionary states, we fitted analytical 
functions to solar-metallicity Parsec models \citep{bressan12} at the terminal 
age main-sequence and the bottom of the red-giant branch. 
Based on these fits we classified stars as giants if:

\begin{equation}
\log(g) <
\begin{cases}
    13.463 - 0.00191\, \teff,& \text{for } \teff > 5000\,K\\
    3.9              & \text{otherwise } 
\end{cases}
\label{equ:class1}
\end{equation}

\noindent
while stars were classified as dwarfs if

\begin{equation}
\log(g) > \frac{1}{4.671} \arctan\left( \frac{\teff-6300}{-67.172}\right) + 3.876  \: .
\label{equ:class2}
\end{equation}

\noindent
Stars were classified as subgiants if they satisfy neither condition (\ref{equ:class1}) 
nor (\ref{equ:class2}). We note that 
the above conditions depend on metallicity and are approximate only. However, they are 
sufficient for qualitatively distinguishing between evolutionary states.
For illustrative purposes only we furthermore distinguish high-luminosity from low-luminosity 
red giants if they have $\logg>2$ (note that strictly speaking this is not a separation 
of evolutionary states, since stars in the red clump are 
more evolved than stars at the tip of the red giant branch).

Figures \ref{fig:rpmcolor}c and \ref{fig:rpmcolor}d show the synthetic 
sample and the control 
sample in a reduced $J$-band proper motion versus $J-K$ color diagram, with 
evolutionary states color-coded according to the \teff\ and \logg\ cuts in the 
top panels. 
The synthetic sample shows that low-luminosity (red) and high-luminosity (blue) red giants
separate cleanly from dwarfs and subgiants, 
while subgiants (green) mostly overlap with the dwarf population (black). 
The control sample shows a similar distribution, with the 
exception of a different color distribution due to the \kep\ target selection 
function. This comparison qualitatively demonstrates that synthetic \textit{Galaxia} 
populations reproduce observations and hence can be used to infer evolutionary states 
that are inaccessible with broadband colors alone.

\subsection{Inference of Stellar Properties using \textit{Galaxia}}

To quantify Figure \ref{fig:rpmcolor}, we inferred stellar properties for 
the synthetic and the control sample using 
$JHKgri$ colors, proper motions, and the \textit{Galaxia} population. 
Given a set of observables  $x=\{J,J-K,H-K,g-r,r-i,\mu\}$
with Gaussian uncertainties $\sigma_x$ and a set of 
intrinsic parameters $y=\{\rm age,\ [Fe/H],\ mass,\ distance\}$, 
the posterior probability of the observed star
having intrinsic parameters $y$ is given by:

\begin{equation}
\begin{split}
p(y|x) \propto p(y)p(x|y) \propto  p(y)\prod_i \exp{\left(-\frac{(x_i-x_i(y))^2}{2 {\sigma_{x,i}^2}}\right)} \: .
\end{split}
\end{equation}

\begin{figure}
\begin{center}
\resizebox{\hsize}{!}{\includegraphics{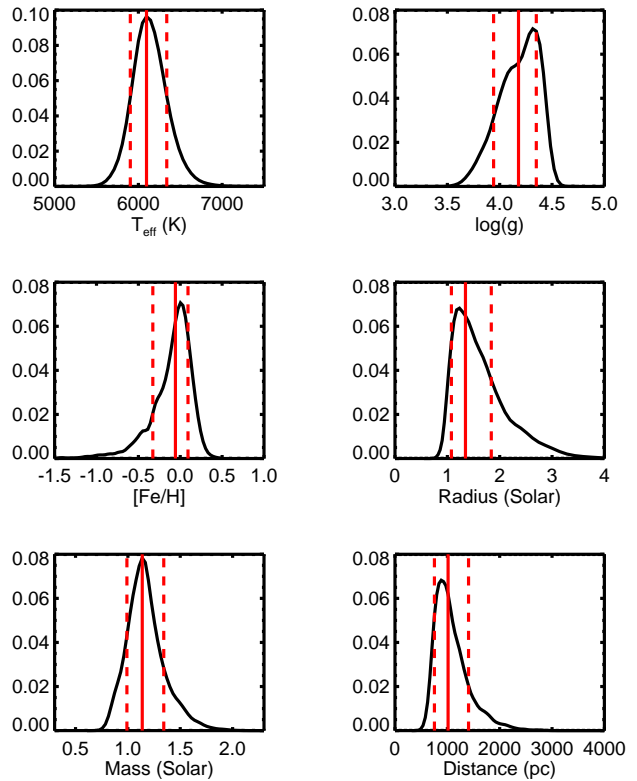}}
\caption{Posterior distributions for \teff, \logg, \feh, radius, mass and distance 
for KIC2438513 ($Kp=13.9, J=12.9$) derived from 
reduced proper motions, $JHKgri$ photometry, and a \textit{Galaxia} synthetic population 
of the \kep\ field. 
Solid and dashed lines show the median and 1-$\sigma$ confidence intervals.}
\label{fig:posteriors}
\end{center}
\end{figure}

Here, $p(y)$  is the prior in the form of an initial mass function, age-metallicity
relation, and age-velocity relation which depend on the
simulated galactic component. We used 
\textit{Galaxia} to generate synthetic stars sampled according to $p(y)$. 
The posterior probability density
function (PDF) was then obtained by summing $p(y_j|x)$
for a given intrinsic parameter $j$ (thereby marginalizing over
all other model parameters), with a stepsize iterated to
yield at least 10 bins within a 1-$\sigma$ confidence interval
around the median.
To avoid discontinuities 
for parameter spaces with a sparse numbers of models (see also next paragraph) 
the distributions were smoothed with a Gaussian with a FHWM of 3 bins.
Note that the volume in age, mass and metallicity covered by each isochrone 
point \citep[which is required to take into account 
different evolutionary speeds, see][]{casagrande11,serenelli13} are 
taken into account through 
the sampling process used to generate the \textit{Galaxia} population. 
Figure \ref{fig:posteriors} shows example posteriors for a star in the \kep\ control sample. 
For each PDF we record the median (solid red line) and 1-$\sigma$ interval around the median 
(dashed red lines). 

\begin{figure*}
\begin{center}
\resizebox{\hsize}{!}{\includegraphics{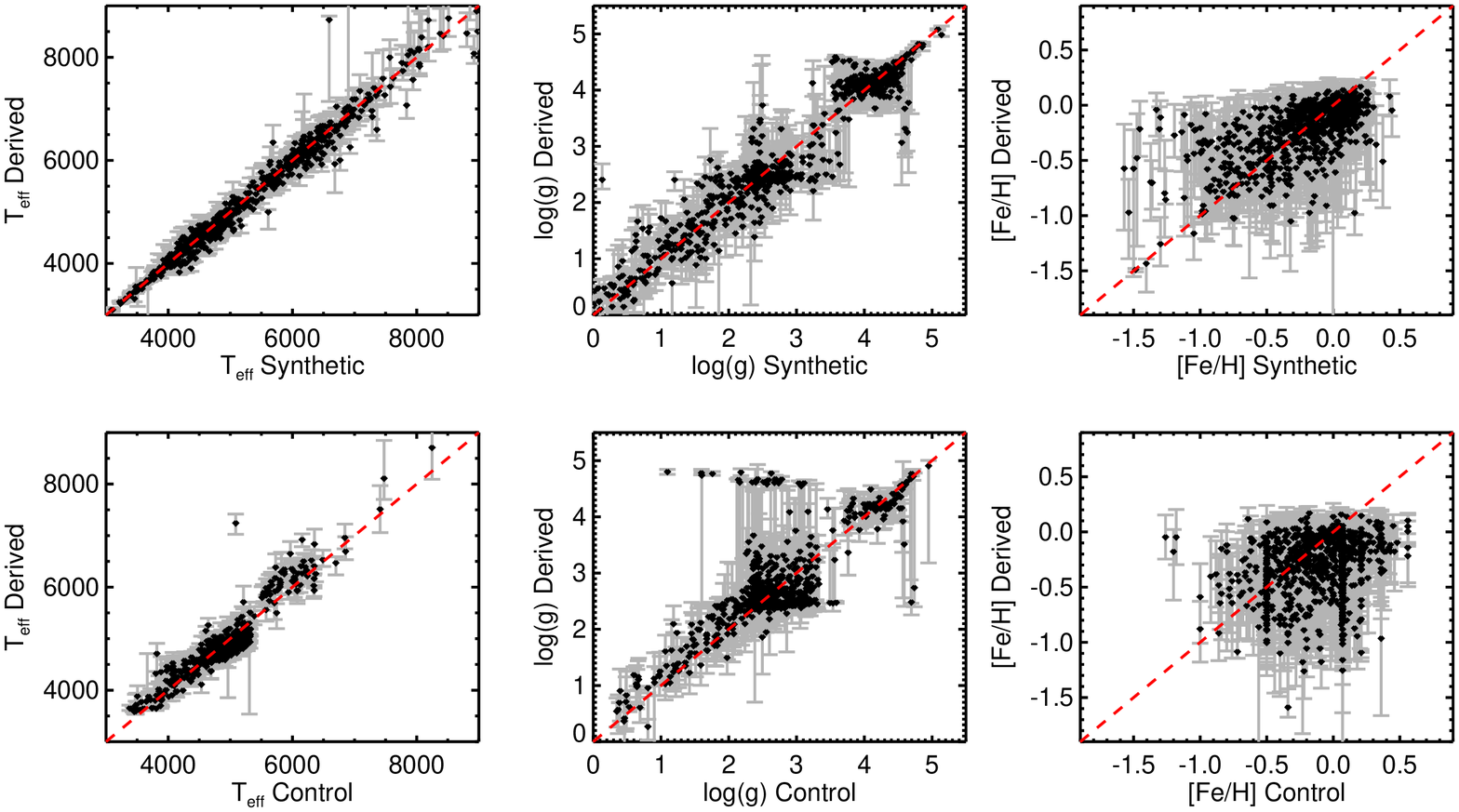}}
\caption{Stellar properties derived from proper motions, colors, and a 
\textit{Galaxia} synthetic population 
versus comparison values for \teff\ (left panels), \logg\ (middle panels) and \feh\ 
(right panels). Comparison values are taken from a synthetic sample of the \kep\ field
(top panels, see also Figure \ref{fig:rpmcolor}a) and the 
\kep\ control sample (bottom panels, see also Figure \ref{fig:rpmcolor}b). Grey lines 
show 1-$\sigma$ error bars and red-dashed line show the 1:1 relation. Only 
5\% of the total sample is shown for clarity.}
\label{fig:testrpm}
\end{center}
\end{figure*}

A key advantage of the method is that \textit{Galaxia} allows
to use proper motions as an observable, which is
not possible for traditional isochrone grids. 
A disadvantage is that in a typical synthetic realization 
by \textit{Galaxia} bright stars are relatively rare, and 
hence parameter inference suffers from small number
statistics. To overcome this we calculated an oversampled 
bright population to ensure that the brightest
0.5\,mag-wide apparent magnitude bin contains at
least 2000 synthetic stars.
We note that for the same reason we
did not make use of spatial information (i.e. galactic latitude
and longitude) of each star, thereby essentially assuming
uniform reddening across a given K2/\kep\ FOV.  
Similarly, we undersampled the faint end of
the synthetic realization to avoid excessive
numbers of synthetic stars.  All subpopulations were calculated to overlap
by at least 0.5\,mag, and are used independently for the 
parameter inference described above (i.e., subpopulations are
not mixed).

To test the method we perturbed observables from the
synthetic sample (Figure \ref{fig:rpmcolor}a) with Gaussian uncertain-
ties typical for the EPIC (0.025\,mag for each photomet-
ric band, and 2.5\,mas/yr for proper motions), and
compared the derived properties to the actual properties 
of each synthetic star. As a second test, we used actual 
data from the \kep\ control sample (Figure \ref{fig:rpmcolor}b)  and compared the 
results to the stellar properties listed in \citet{huber14} (all of which were derived 
from asteroseismology and/or spectroscopy). 

The results are shown in Figure \ref{fig:testrpm}. As expected
the synthetic sample performs better than the control
sample due to differences between the real data and
the synthetic data. \teff\ is well recovered, 
with the control sample showing a slight bias, being on average $\approx$\,80\,K hotter than the comparison
values. We 
attribute this bias to differences in the color-temperature scale and/or extinction maps 
between \textit{Galaxia} and \citet{huber14}, which mostly adopted \teff\ values from 
\citet{pinsonneault11}. The right panels of Figure \ref{fig:testrpm} show 
that the method has very little sensitivity to \feh, and essentially recovers 
the \feh\ prior from synthetic population. \textit{Therefore, 
metallicities derived in this work are valid in a statistical sense only}.

The \logg\ comparison (middle panels of Figure \ref{fig:testrpm}) 
shows that 1--4\% of giants are 
misclassified as dwarfs, while 4--7\% of dwarfs are misclassified as giants. Closer 
inspection revealed that giants classified as dwarfs are evolved stars with 
unusually high proper motions, some of which are likely due to catalog errors. 
The \logg\ comparison for the \kep\ sample 
shows a slight systematic bias for giants, with derived \logg\ values being 
systematically larger. This is consistent with the \teff\ bias 
discussed above, because in the absence of a strong \logg\ constraint an 
overestimated temperature on the red-giant branch yields an overestimated \logg.
As expected from Figure \ref{fig:testrpm}, the dwarf-subgiant separation is 
considerably less successful than for dwarfs and giants: 
56--72\% of subgiants are classified as 
dwarfs, while 9\% of dwarfs are misclassified as subgiants. 
We emphasize, however, that this is in most cases appropriately captured within the 
uncertainties: $\approx$\,79\% (for the control sample) and 
$\approx$\,93\% (for the synthetic sample) of all true subgiants would be classified 
as subgiants within their respective 1-$\sigma$ error bars in \teff\ and \logg.

We conclude that proper motions and colors from the EPIC combined with a synthetic 
population generated by \textit{Galaxia} 
yield stellar classifications and uncertainties which are suitable for 
large populations of stars. The overall residual scatter from the tests performed above is 
$\approx 2-3$\% for \teff, $\approx 0.3$\,dex in \logg\, and $\approx 0.3$\,dex in \feh. This is 
consistent with the performance of the KIC \citep{bruntt10,kallinger10,verner11,huber14},
despite the lack of Sloan $z$ and $D51$ photometry in the EPIC.

\subsection{Stellar Populations in C1--C8}

\begin{figure*}
\begin{center}
\resizebox{\hsize}{!}{\includegraphics{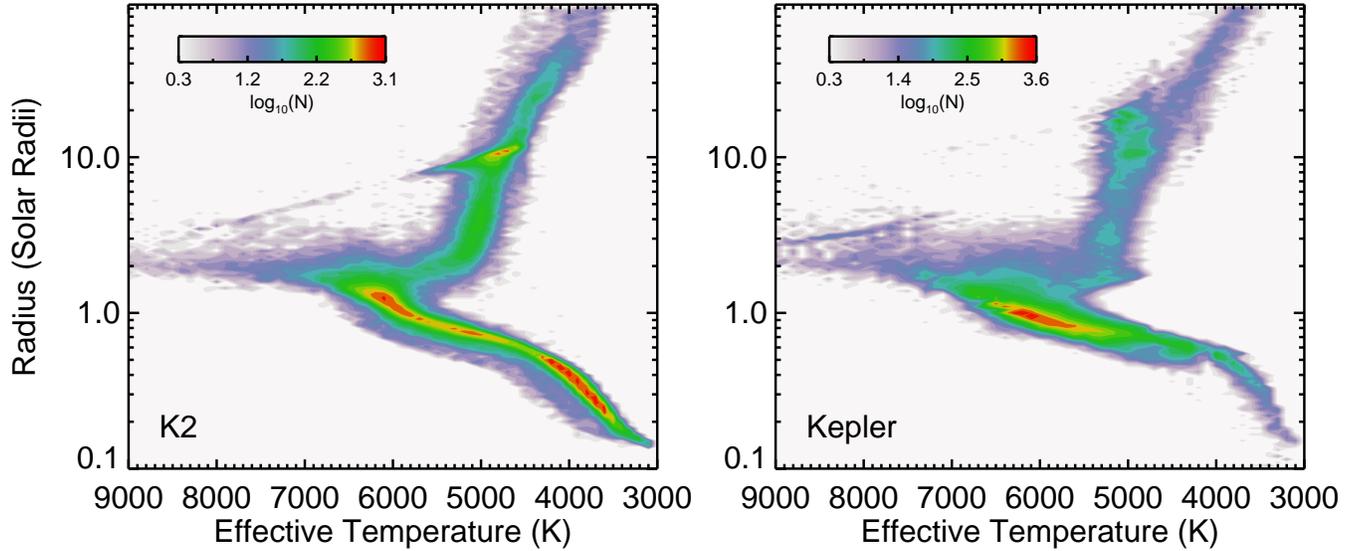}}
\caption{Stellar radius versus effective temperature for \nclass\ K2 targets classified 
in this work (left) and $\approx$\,190,000 \kep\ targets taken from \citet{huber14}. Colors 
show the logarithmic number density. Note that differences in the 
shape of the distribution result mostly from the different isochrones (Padova 
for K2, Dartmouth for \textit{Kepler}) which were applied for the 
classification.}
\label{fig:hrd}
\end{center}
\end{figure*}

To classify K2 targets, we calculated \textit{Galaxia} populations for each K2 field and 
applied the method described in the previous section to all 
EPIC IDs selected for observations. K2 targets were extracted from the 
K2 web page at the Kepler Science 
Center\footnote{\url{http://keplerscience.arc.nasa.gov/index.html}}.
In addition to colors and proper motions, 
we used parallaxes from Hipparcos \citep{vanleeuwen07b} and spectroscopic 
information (\teff, \logg\ and \feh) from the RAVE DR4 \citep{kordopatis13}, 
LAMOST DR1 \citep{luo15} and APOGEE DR12 \citep{alam15} surveys. The spectroscopic sources were 
matched to the EPIC coordinates with a radius of 3''.
The observables $x$ used to classify a given target were then chosen according to 
the following priority order:

\begin{itemize}
\item[1)] $x=\left\{J,J-H,H-K,g-r,r-i,\pi\right\}$ if Hipparcos parallaxes ($\pi$) were 
available. If no $JHK$ photometry was available for bright stars with 
parallaxes, $x=\left\{V,B-V,\pi\right\}$ 
were used. If only a small number of models were available within the uncertainties 
(e.g.\ very close M dwarfs) we fitted absolute magnitudes instead of parallaxes 
and apparent magnitudes, thereby removing the distance constraint. For these cases 
no $E(B-V)$ values are reported.

\item[2)] $x=\left\{J,\teff,\logg,\feh\right\}$ if a RAVE, LAMOST or APOGEE 
classification were available. We assumed default uncertainties of $150$\,K in \teff,
$0.15$\,dex in \logg\ and $0.15$\,dex in \feh\ for all spectroscopic 
classifications. We assumed solar metallicity if no metallicity estimate was 
provided by APOGEE.

\item[3)] $x=\left\{J,J-K,H-K,g-r,r-i,\mu\right\}$ if proper motions were 
available.

\item[4)] $x=\left\{J,J-K,H-K,g-r,r-i \right\}$ if only colors were available.
\end{itemize}

\noindent
We assumed a minimum uncertainty of 1\,mas/yr for all total proper motions and 
0.03\,mag for all colors.
K2 targets which are either flagged as extended objects (e.g. galaxies) or 
have neither valid 2MASS photometry nor a Hipparcos parallax in the EPIC are not classified. 
For C1--C8 this affects $\approx$\,12\% of all targets.

Figure \ref{fig:hrd} shows a \teff-radius number density distribution of the 
\nclass\ K2 targets classified in this work compared to $\approx$\,190,000 targets 
in the \kep\ field. Stellar properties for the \kep\ sample were 
taken from \citet{huber14}. The distributions demonstrate the 
preferential selection of K/M dwarfs for K2, 
which were underrepresented in the \kep\ sample compared to the larger 
fraction of sun-like dwarfs. K2 also includes a larger relative fraction 
of red giants compared to dwarfs, a consequence of the K2 galactic archeology program 
to map the age and metallicity distributions of stars 
using asteroseismology \citep[e.g.,][]{miglio12,casagrande14,apokasc,stello15}.
Note that the diagonal ``band'' near $\teff\approx$\,6500--7000\,K and 
$R\approx$\,4--7\rsun\ is due 
to the sparseness of metal-poor horizontal branch models in \textit{Galaxia}.

\begin{figure*}
\begin{center}
\resizebox{\hsize}{!}{\includegraphics{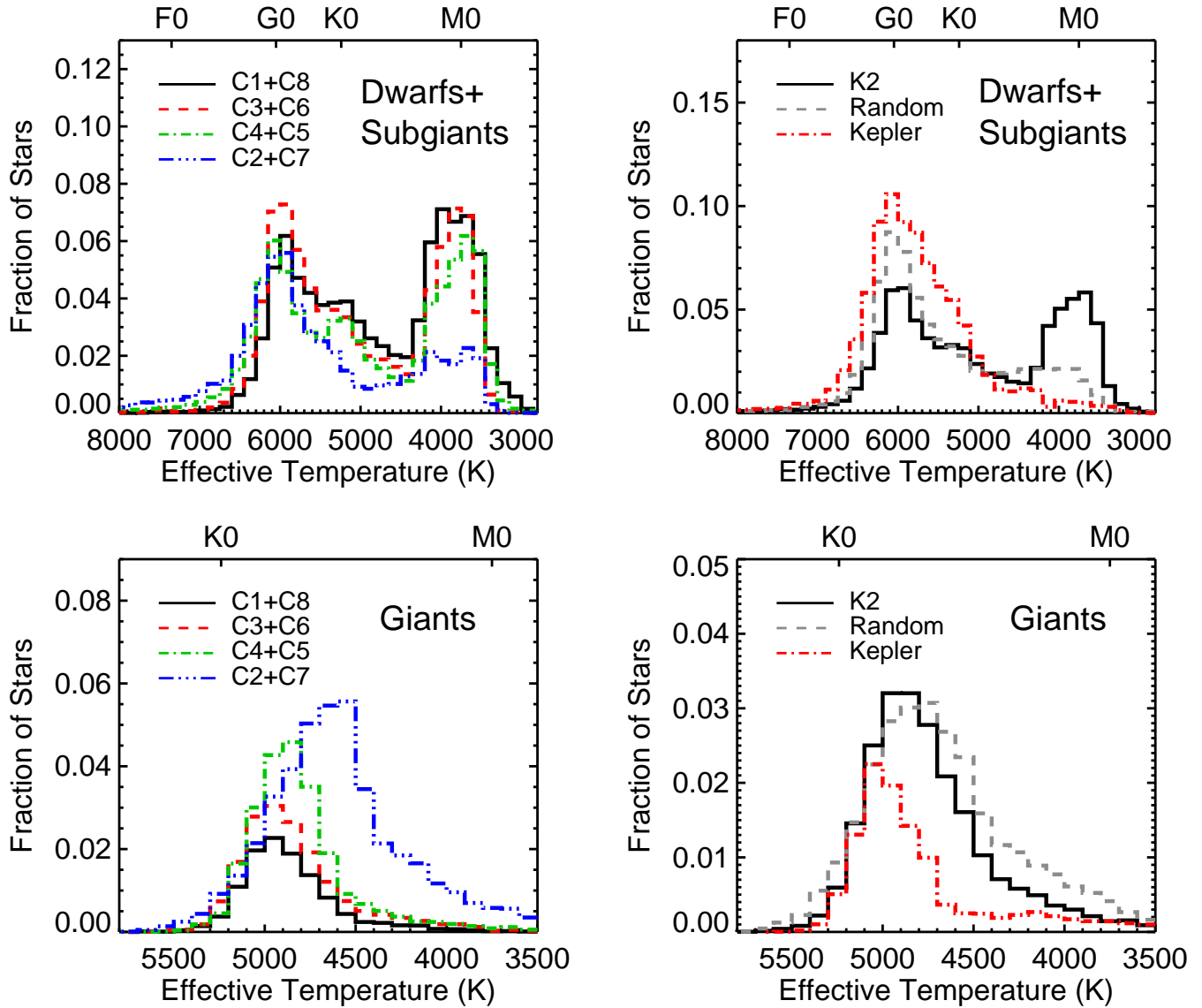}}
\caption{Left panels: Effective temperature distributions for K2 targets classified as 
subgiants and dwarfs (top) and giants (bottom). Targets have been grouped
into campaigns with similar heights above the galactic plane: 
$|b|\approx 57$--$58^{\circ}$ (C1+C8),
$|b|\approx 50$--$52^{\circ}$ (C3+C6),
$|b|\approx 26$--$32^{\circ}$ (C4+C5),
$|b|\approx 15$--$19^{\circ}$ (C2+C7).
The top axis labels show representative spectral types (see also Table \ref{tab:class}).
Right panels: Same as the left panels but for all K2 campaign targets, 
the \kep\ sample, and a random sample 
drawn from \textit{Galaxia} synthetic populations which match the apparent magnitude 
distribution of K2 targets.}
\label{fig:classification}
\end{center}
\end{figure*}

Figure \ref{fig:hrd} shows systematic differences between the K2 and \kep\ samples 
due to the use of different models: 
\textit{Galaxia} adopts Padova isochrones \citep{girardi00,marigo07,marigo08} while the 
\kep\ stellar properties catalog is based on isochrones from the Dartmouth Stellar 
Evolution Program \citep{dotter08}. Specifically, radii for M dwarfs at fixed \teff\ 
are systematically smaller in Padova models, which disagrees with empirical measurements 
from long-baseline interferometry \citep{boyajian12b,boyajian13} or low-mass eclipsing 
binary systems \citep{kraus11,carter11}. This discrepancy has been improved in the 
updated Parsec isochrones \citep{bressan12}, which however have not yet been implemented 
in \textit{Galaxia}. Hence, users should be aware that radii for M dwarfs derived in this work 
can be underestimated by up to $\approx$20\%. On the other hand, classifications of 
giants near the red clump are more 
accurate for the K2 sample because the Dartmouth models adopted by \citet{huber14} did not 
include He-core burning stars. This can be seen by the correct 
position of the red clump near $\approx10\rsun$ for the K2 sample in Figure \ref{fig:hrd}.

To quantify the classifications of C1--C8 targets, Figure \ref{fig:classification} 
shows the effective temperature distribution for dwarfs and subgiants 
(top panels) and giants (bottom panels)
subdivided into campaigns at similar heights above the galactic plane (left panels). 
The luminosity classification was performed using the conditions in 
Equations (\ref{equ:class1}) and (\ref{equ:class2}). 
The distributions show several interesting differences 
in the target populations between different campaigns. For example, campaigns which 
are close to the galactic plane (C2 and C7) show a much 
lower fraction of M dwarfs but a larger relative fraction of hot dwarfs and cool giants. 
We suspect that this is partially due to the increased number of giants in a 
magnitude limited sample in the galactic plane as well as the effect of higher 
interstellar reddening, the combination of which led to a 
preferential selection of cool giants and hot dwarfs compared to cool dwarfs.
 The effect can also be seen in the temperature 
distribution of giant stars (bottom left panel): 
since more red stars are classified as giants in lower latitude fields, the average 
\teff\ for giants is cooler than in higher latitude fields.
We caution, however, that uncertainties in the reddening model 
adopted in \textit{Galaxia} may also contribute to this effect. The top left panel 
furthermore shows a slight overabundance of K dwarfs and subgiants with $\teff\approx5200$\,K 
in fields at high latitudes, which can also tentatively be seen in Figure \ref{fig:hrd}. 
The reason for this overabundance is unknown, but is likely due to a systematic 
misclassification of giants as dwarfs rather than a preferential target 
selection by the community. 

The right panels of 
Figure \ref{fig:classification} compare the \teff\ distribution of K2 targets to 
the \kep\ sample \citep{huber14}. Additionally, we show a randomly drawn sample from the 
synthetic populations generated by \textit{Galaxia}, 
which for each campaign reproduces the apparent magnitude distribution of 
the actual K2 target list. 
As expected, for dwarfs and subgiants there is a clear overabundance of K/M dwarfs for K2 
compared to \kep\ and the random sample, while the \kep\ sample is biased 
towards sun-like dwarfs and against cool dwarfs. For giants the distribution
is largely consistent with a random selection, in agreement with the target selection function 
of the galactic archeology program (which uses a $J-K>0.5$ color cut).
Overall, the distributions are consistent with the primary K2 science programs. 

\begin{table*}
\begin{center}
\caption{Stellar classifications of K2 Targets}
\begin{tabular}{c r r r r r r r r r r r}
\hline
Spectral Type & C1 & C2 & C3 & C4 & C5 & C6 & C7 & C8 & C1--C8 & Random & \kep\ \\
\hline
\multicolumn{12}{c}{Dwarfs \& Subgiants (\%)} \\
  A  &     0.1  &     2.7  &     0.5  &     1.9  &     0.5  &     0.4  &     3.4  &     0.2    &     1.0  &     1.0  &     1.9   \\
  F  &     5.9  &    22.0  &    12.6  &    18.3  &    12.9  &    11.6  &    12.0  &     9.8    &    12.7  &    19.9  &    30.7   \\
  G  &    14.5  &    17.3  &    25.7  &    17.8  &    20.4  &    28.3  &    20.6  &    35.0    &    23.0  &    27.6  &    40.4   \\
  K  &    44.5  &     9.9  &    34.0  &    22.5  &    29.7  &    30.4  &    17.0  &    33.4    &    28.9  &    21.1  &    14.4   \\
  M  &    22.5  &     4.1  &    11.4  &    16.2  &    13.3  &    11.4  &     5.6  &    11.1    &    12.4  &     3.1  &     1.4   \\
\hline
\multicolumn{12}{c}{Giants (\%)} \\
  G  &     0.3  &     0.9  &     0.5  &     0.6  &     0.3  &     0.8  &     1.9  &     0.5    &     0.7  &     1.4  &     0.4   \\
  K  &    12.1  &    41.4  &    15.3  &    22.2  &    22.7  &    17.1  &    37.4  &     9.9    &    20.9  &    24.9  &    10.4   \\
  M  &     0.0  &     1.6  &     0.1  &     0.4  &     0.2  &     0.1  &     2.2  &     0.1    &     0.5  &     0.9  &     0.6   \\
\hline
\multicolumn{12}{c}{Total Classified (\%)} \\
     &    83.9  &    98.0  &    83.1  &    95.1  &    87.5  &    86.5  &    94.7  &    80.2  &    87.5  & -- & --  \\
\hline
\end{tabular} 
\label{tab:class}
\flushleft Notes: Luminosity classifications were assigned from the derived \teff\ and \logg\ 
using the criterium in Equation \ref{equ:class1}. Spectral types were assigned 
using the following temperature 
ranges: $>7350$\,K (A), $6050-7350$\,K (F), $5240-6050$\,K (G), 
$3750-5240$\,K (K) and $<3750$\,K (M). The second to last column lists classifications 
for a random sample drawn from the \textit{Galaxia} simulations with the same apparent magnitude 
distribution as the K2 sample. The classifications of \kep\ targets in the last column 
is based on \citet{huber14}.
\end{center}
\end{table*}

Table \ref{tab:class} summarizes the spectral types of K2 targets in 
C1--C8 presented here. 
Overall, the most common K2 targets are K--M dwarfs ($\approx$\,\nkmdwarfs\% of 
all classified targets) followed by F--G dwarfs ($\approx$\,\nfgdwarfs\%) 
and red giants ($\approx$\,\nkgiants\%). We emphasize that there is 
considerable variation in these classifications from campaign to campaign.

\subsection{K2 Exoplanet Host Stars}

\begin{figure}
\begin{center}
\resizebox{\hsize}{!}{\includegraphics{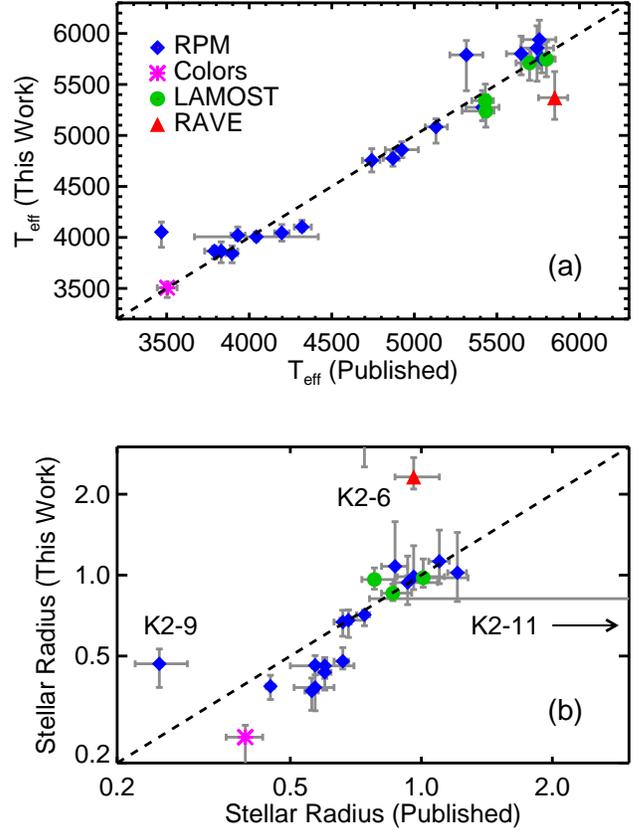}}
\caption{Comparison of effective temperatures (panel a) and radii (panel b) of K2 
planet host stars derived in this work versus values published in the literature. 
Different colors and symbols indicate the classification method used. 
Three host stars which are discussed in more detail in the text are marked.}
\label{fig:hosts}
\end{center}
\end{figure}

Figure \ref{fig:hosts} compares effective temperatures (panel a) and stellar 
radii (panel b) from this work compared to values published in K2 planet 
discovery papers. Most literature values are taken from \citet{montet15}, 
who derived posteriors by comparing broadband colors to 
Dartmouth isochrones using Nested Sampling \citep{morton15}. 
Additionally we have added K2-21 \citep{petigura15}, 
K2-22 \citep{sanchis15}, as well as host stars from \citet{sinukoff15}, with 
stellar properties derived from high-resolution spectra using Specmatch \citep{petigura15b}.
We note that K2-2 \citep{vanderburg15} was not included 
since it was observed during the K2 engineering run, which is not covered in our 
catalog.

Figure \ref{fig:hosts}a shows good agreement for \teff, with an offset of $\approx$\,1\% and a 
residual scatter of $\approx$\,3\%, which is consistent within the reported uncertainties. 
Figure \ref{fig:hosts}b shows that radii for stars below $\lesssim0.6\rsun$ are 
systematically lower in our sample, which is consistent with the systematic offset 
between Padova and Dartmouth isochrones discussed in the previous Section. Radii for 
G--K type dwarfs are broadly in agreement. 

Several K2 planet hosts show significant radius differences which cannot 
be explained by isochrone systematics. K2-11 (EPIC\,201596316) yielded a bimodal dwarf-giant 
classification in \citet{montet15} ($\teff=5433^{+49}_{-144}$\,K,$R=5.15^{+0.20}_{-4.39}\,\rsun$), 
whereas our results (based on a spectroscopic classification from LAMOST) favor a K dwarf 
($\teff=5240^{+160}_{-160}$\,K, $R=0.82^{+0.05}_{-0.05}\rsun$), which 
would firmly place the planet in Earth-sized regime ($\approx1.2\rearth$). Conversely, 
K2-6 (EPIC\,201384232) was classified as a sun-like dwarf by \citet{montet15} 
($\teff=5850^{+79}_{-98}$\,K, $R=0.96^{+0.14}_{-0.09}\rsun$), consistent with high-resolution
spectroscopy by \citet{vanderburg15b}, while our results (based on a 
spectroscopic classification from RAVE) favor a 
K-subgiant ($\teff=5370^{+255}_{-213}$\,K, $R=2.32^{+0.42}_{-0.23}\rsun$). 
Our results for K2-9 (EPIC\,201465501) favor a hotter and larger late-K dwarf 
($\teff=4052^{+99}_{-148}$\,K, $R=0.47^{+0.06}_{-0.08}\rsun$) compared to the mid-M classification by 
\citet{montet15} ($\teff=3468^{+20}_{-19}$\,K, $R=0.25^{+0.04}_{-0.03}\rsun$), 
which was later 
confirmed spectroscopically by \citet{schlieder16}. Closer inspection 
showed that while $J-K$ is compatible with a mid-M dwarf, $H-K$ indicates a hotter 
solution which combined with the fact that very red $J=12.5$\,mag dwarfs should be rare 
skews our posteriors towards $\teff \approx 4000$\,K. 
Finally, EPIC\,201713348 and EPIC\,203826436 are classified as evolved stars ($R>3\rsun$) 
while the spectroscopic classifications by \citet{sinukoff15} imply late-type dwarfs 
($R<1\rsun$). We suspect that the discrepancies are due to underestimated uncertainties in the 
proper motions adopted in the EPIC (which are consistent with evolved stars), and 
note that the spectroscopic solutions should be preferred.

Finally, we compared \teff\ and \logg\ values for 36 planet-candidate hosts with the catalog by 
\citet{vanderburg15b}, who presented high-resolution spectra analyzed using 
SPC \citep{buchhave12,buchhave14}.
We find satisfactory agreement with residual scatter of $\approx$\,180\,K in \teff\ 
and $\approx$\,0.2\,dex for \logg. 
In addition to the two hosts discussed above, seven stars are 
classified as evolved ($\logg \approx$ 2.5--4.0) in the EPIC 
while they are classified as main-sequence stars ($\logg>4.3$) based on SPC. 
Three of these hosts (EPIC\,201384232, EPIC\,205944181 and EPIC\,205950854) 
are based on RAVE 
spectroscopic classifications, while the remaining four (EPIC\,204890128, EPIC\,206011496, 
EPIC\,206026904 and EPIC\,206114630) were classified using proper motions. While 
the catalog uncertainties should in most cases capture such misclassifications, these differences 
illustrate that stellar properties in the EPIC are no substitute for follow-up 
observations to precisely characterize planets discovered by K2.

\subsection{K2 Asteroseismic Giants}
\label{sec:rgs}

Stellar oscillations can be used to precisely measure fundamental 
properties in red giants with long-cadence data \citep[e.g.][]{kallinger10} and hence 
test derived stellar properties in the catalog. Since the majority of stars 
with $\logg\approx2-3.4$ and $Kp<15$ should show detectable oscillations 
with K2 \citep{stello15}, asteroseismic analyses are
also efficient to discern luminosity classes of \kep\ and K2 targets \citep{huber13}.

We have performed an asteroseismic 
analysis of 678 K2 targets in C5 which have been targeted by the
SAGA survey \citep{casagrande14}. We used PDC-SAP light curves and 
applied iterative 4-$\sigma$ clipping as well as a high-pass filter 
with a 2-day width to prepare the data for the asteroseismic analysis. A second, 
independent analysis used gap-filling to minimize the effects of the 6 hour 
repointing cycle \citep{stello15}. Both datasets were searched for oscillations using
the method by \citet{huber09}. 

\begin{figure}
\begin{center}
\resizebox{\hsize}{!}{\includegraphics{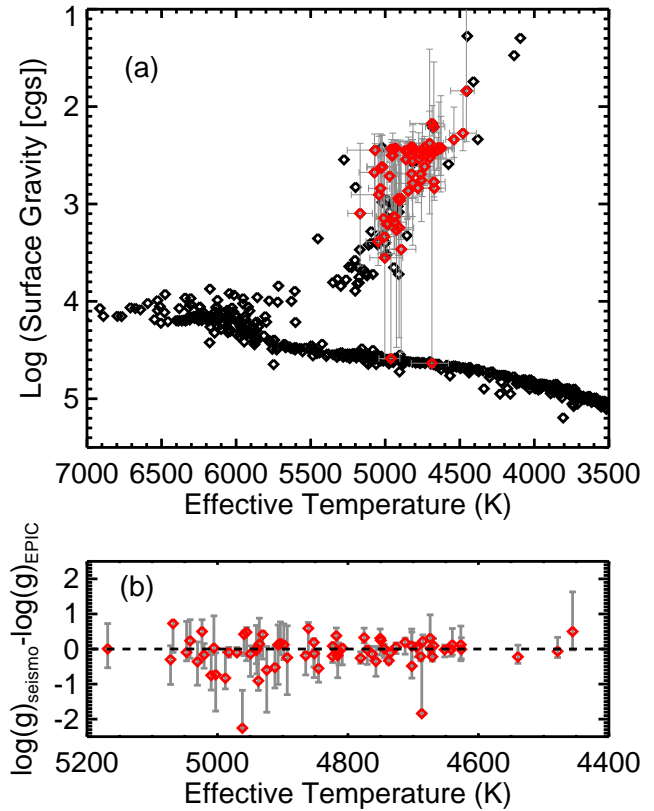}}
\caption{(a) Surface gravity versus effective temperature taken from the EPIC 
for 612 K2 targets observed in C5. Red triangles with error bars show asteroseismically 
confirmed giants based on the analysis of K2 data (see Section \ref{sec:rgs}). 
(b) Difference between \logg\ in the EPIC and \logg\ derived from asteroseismology. 
The residual scatter is $\approx 0.35$\,dex.}
\label{fig:seismo}
\end{center}
\end{figure}

Out of the 678 targets that were analyzed, 612 have been classified in the EPIC.
Figure \ref{fig:seismo}a shows a \logg-\teff\ diagram using the 
EPIC classifications for all 612 stars (black symbols) and the subset of 70 stars for which 
oscillations were detected (red symbols). Of the stars with detected oscillations 
two were misclassified as dwarfs in the EPIC, confirming the low misclassification 
rate derived in Section 5.3. We note that both misclassified giants have 
uncertainties that place them on the giant branch within 1-$\sigma$.
Twenty seven stars are predicted to be red giants by the EPIC, but do not show 
oscillations. Inspection of the K2 data showed that about 4 stars oscillate 
with periods too long to be resolved with a single K2 campaign (high-luminosity giants), 
while 18 stars are low S/N detections or targets for which the light curves are heavily 
contaminated by the 6 hour pointing artefacts. The remaining 5 stars are likely dwarfs that 
are misclassified as giants due to erroneous proper motions.

To test the accuracy of the derived stellar properties we combined the 
measured frequency of maximum power ($\numax$) for each oscillating giant with \teff\ 
from the EPIC to calculate \logg, and compared these values to those in the EPIC 
in Figure \ref{fig:seismo}b. The residuals shows a scatter of $\approx$\,0.35\,dex with 
no systematic offset, confirming that the EPIC classifications do not only 
discern dwarfs from giants with a low false positive rate, 
but also have some sensitivity to evolutionary states on the red giant branch.

\subsection{K2 Stellar Properties Catalog and Shortcomings}

Table \ref{tab:catalog} lists stellar properties, distances and extinction values 
for the \nclass\ K2 targets in C1--C8 (88\% of the total target sample)
derived in this work. Reported values are the posterior median and 1-\,$\sigma$ confidence 
interval around the median. 
Note that for a small fraction ($\approx$0.2\%) of stars with bimodal distributions 
where the median would result in unphysical stellar properties we adopted the 
best-fit and 1-\,$\sigma$ confidence interval around the best fit instead. 

The table also includes a flag denoting which observables were used to derive the 
listed properties. From all \nclass\ classified C1--8 targets, 1\% were classified 
using parallaxes and colors, 7\% using spectroscopy, 81\% using reduced proper motions 
and colors, and 11\% using colors only.
The source code used to derive the classifications is 
available\footnote{\url{https://github.com/danxhuber/galclassify}} and can be used 
to extract posterior distributions for each target.

\begin{table*}
\begin{footnotesize}
\begin{center}
\caption{Stellar Properties of K2 Targets}
\begin{tabular}{c c c c c c c c c c}
\hline
EPIC & $\teff$ & $\logg$ & $\feh$ & $R$ & $M$ & $\rho$ & d & E(B-V) & Flag \\
     &     (K)    &   (cgs)   &     & (\rsun) & (\msun) & (\rhosun) & (pc) & (mag) & \\
\hline
\input{table5.tex} 
\hline
\end{tabular}
\label{tab:catalog}
\flushleft Notes: Reported values are the posterior median and 1-$\sigma$ confidence 
interval around the median. Flags indicate which observables were used to derive 
the classification: 
hip = Hipparcos parallax + colors;
apo = APOGEE spectroscopic classification; 
rav = RAVE spectroscopic classification; 
lam = LAMOST spectroscopic classification;
rpm = reduced proper motions + colors; 
col = colors only;
\end{center}
\flushleft
\end{footnotesize}
\end{table*}

We stress that while the catalog should be useful for a 
first-look classification of K2 targets, care should be taken when using the 
catalog for scientific investigations. 
Specifically, the following shortcomings should be expected:

\begin{itemize}
\item A significant fraction ($\approx$\,55-70\%) of subgiants are misclassified as dwarfs, 
and hence their radii will be systematically underestimated. This uncertainty will in 
most cases be captured in the posterior distributions (and hence the quoted 
confidence intervals), but using point estimates for a large population may lead to 
biased results.

\item By construction, stars which are rare in the \textit{Galaxia} populations
(such as pre-main sequence stars, massive main-sequence stars, AGB stars, and white dwarfs) 
will either be missing or misclassified in the catalog. Users are encouraged 
to use the observational information in the EPIC (such as colors) to identify and 
characterize such sources.

\item Metallicities for stars without spectroscopic input 
are statistical estimates only and should not be used for scientific investigations.

\item The reddening model in \textit{Galaxia} does not take into account low 
extinction regions in the solar neighborhood (the ``local bubble''), and hence 
$E(B-V)$ values are likely overestimated for stars within $\lesssim$\,100\,pc (in 
particular for campaigns close to the galactic plane).

\item Uncertainties provided in the catalog do not account for  
systematic errors in isochrone models such as different physical prescriptions of 
mixing length, convective core-overshooting, helium abundances, and mass loss. We 
have adopted the following lower limits for uncertainties: 
1\% in \teff, 0.02\,dex in \logg, 
0.05\,dex in \feh, 2\% in radius and mass, 3\% in distance, and 6\% in density.

\item \textit{Galaxia} uses Padova isochrones to predict masses and radii, which
show systematic differences to other model grids. In particular, Padova models 
are known to systematically underpredict radii for cool 
dwarfs by up to 20\% \citep{boyajian12}. Hence, users are encouraged to apply 
empirical corrections to radii of cool dwarfs, or use the derived \teff\ values with 
empirical calibrations to estimate radii \citep{mann15}. Future revisions of \textit{Galaxia} 
will implement improved isochrones to alleviate this problem. 
\end{itemize}

\section{Summary \& Conclusions}

We presented the Ecliptic Plane Input Catalog (EPIC) for the K2 mission, 
which serves a similar purpose as the Kepler Input Catalog (KIC) for the \kep\ Mission. 
The EPIC is a federation of the Hipparcos, Tycho-2, 2MASS, 
UCAC4 and SDSS catalogs and contains EPIC identifiers, coordinates, \kep\ magnitudes, 
photometry, astrometry and cross-matched identifiers for each source.  
\kep\ magnitudes in the EPIC are accurate to $\approx 0.1$\,mag, except if they are 
based on $J$-band photometry only for which discrepancies of 0.5\,mag or more can 
be expected. The EPIC is magnitude complete down to $Kp\approx17$ for 
typical campaigns, and down to $Kp\approx19$ covered by SDSS. For non-SDSS fields the 
completeness is set by 2MASS, and hence the catalog is expected to be significantly 
more incomplete for blue objects. 
We furthermore described the procedures to 
mitigate missing sources which are proposed for K2 observations, and discussed 
several known shortcomings (such as duplicate sources and incompleteness) 
which should be kept in mind when using the EPIC. At the time of writing, the EPIC includes 
48 million sources for C0--C13. 

Unlike the KIC, the EPIC does not include stellar classifications for each 
catalogued source. To support the K2 community we have derived 
stellar properties (temperatures, surface gravities, metallicities, 
radii, masses, densities, distances, and extinctions) for \nclass\ K2 targets in C1--8
using colors, proper motions, spectroscopy, parallaxes 
and stellar population models for each field. We demonstrated 
that these classifications allow a reliable ($\approx 95\%$ success rate) separation between 
giants and dwarfs, but suffer from biases to classify $\approx 55-70$\% of 
subgiants as dwarfs. However, this bias should in most cases be 
appropriately captured in the provided uncertainties. Our tests imply a 
precision of $\approx 2-3$\% for \teff, $\approx 0.3$\,dex in \logg\, and $\approx 0.3$\,dex 
in \feh, which is consistent with  typical formal uncertainties for G-type stars 
in the catalog: 140\,K in \teff, $\approx 0.3$\,dex in \logg, $\approx 0.3$\,dex 
in \feh, $\approx 40$\% in radius, $\approx 10$\% in mass, $\approx 60$\% in density, and 
$\approx 40$\% in distance. We emphasize that the uncertainties vary considerably 
with spectral type and evolutionary state, and do not take into account systematic 
errors between different isochrones.

We showed that the K2 target sample in C1--8 is dominated by
K--M dwarfs (\nkmdwarfs\% of all classified targets) followed by F--G dwarfs 
(\nfgdwarfs\%) and K giants (\nkgiants\%), consistent  
with key K2 science programs to detect transiting exoplanets and conduct galactic archeology 
using oscillating red giants. We showed that these distributions vary considerably over 
different campaigns: 
low galactic latitude fields appear to have a larger relative fraction of giants, presumably 
as a result of the increased giant contamination and reddening introducing a bias in 
the target selection.
We used a set of asteroseismic giants to verify the giant-dwarf classifications in the 
catalog but also find significant differences to spectroscopic classifications of several K2 
exoplanet host stars, which emphasize the need for detailed 
spectroscopic follow-up to reliably characterize detected transiting planets.
Future observations from large-scale spectroscopic surveys such as 
APOGEE \citep{allende08} and GALAH \citep{desilva15} as well as parallaxes from Gaia \citep{perryman05}
will considerably improve the classifications provided here.

The EPIC is hosted at the Mikulski 
Archive for Space Telescopes 
(MAST)\footnote{\url{http://archive.stsci.edu/k2/epic/search.php}} 
and EPIC information on K2 targets (including stellar classifications)
and confirmed planet hosts can be found at the 
NASA Exoplanet Archive\footnote{\url{http://exoplanetarchive.ipac.caltech.edu/}}. 
The source codes to produce an EPIC\footnote{\url{https://github.com/danxhuber/k2epic}}
and to derive stellar classifications (including posteriors) 
for K2 targets\footnote{\url{https://github.com/danxhuber/galclassify}} are 
publicly available.
We note that the EPIC is 
continuously updated for new K2 campaign fields, and users are encouraged to 
regularly check the online documentation at 
MAST\footnote{\url{http://archive.stsci.edu/k2/epic.pdf}} 
for updates.

\section*{Acknowledgments}
We thank Tim Bedding, Jeff van Cleve and Sebastien Lepine for comments and discussions, and 
numerous members of the K2 community for valuable feedback and bug reports. We also thank 
Randy Thompson, Scott Fleming and everyone at MAST for their help in making the EPIC available to 
the community. We are furthermore grateful for the support of the  Kavli Institute for Theoretical Physics
in Santa Barbara, where part of this research was conducted.

D.H. acknowledges support by the Australian Research Council's Discovery Projects 
funding scheme (project number DE140101364) and support by the National Aeronautics 
and Space Administration under Grant NNX14AB92G issued through the Kepler 
Participating Scientist Program.
Funding for the \kep\ Mission is provided by NASA's Science Mission Directorate. 
This publication makes use of data products from the Two Micron All Sky Survey, which is a joint project of 
the University of Massachusetts and the Infrared Processing and Analysis Center/California Institute of Technology, 
funded by the National Aeronautics and Space Administration and the National Science Foundation.
This research made use of the AAVSO Photometric All-Sky Survey (APASS), funded by the Robert Martin Ayers 
Sciences Fund. The Second Palomar Observatory Sky Survey (POSS-II) was made by the California Institute of 
Technology with funds from the National Science Foundation, the National Geographic Society, the Sloan 
Foundation, the Samuel Oschin Foundation, and the Eastman Kodak Corporation. 
This research was supported in part by the National Science Foundation under Grant No. NSF PHY11-25915.

\newpage
\bibliographystyle{aasjournal}
\bibliography{/Users/daniel/Dropbox/tex/references}

\end{document}

%% file: table5.tex
      201425819  &         $3937^{+117}_{-70}$ &   $4.849^{+0.042}_{-0.035}$ &  $-0.024^{+0.120}_{-0.210}$ &   $0.431^{+0.035}_{-0.051}$ &   $0.486^{+0.053}_{-0.053}$ &                $5.79^{+1.47}_{-0.86}$ &           $202^{+23}_{-25}$ &   $0.032^{+0.014}_{-0.014}$ &     rpm \\
      201426057  &        $3935^{+231}_{-192}$ &   $4.864^{+0.105}_{-0.075}$ &  $-0.113^{+0.200}_{-0.240}$ &   $0.411^{+0.079}_{-0.100}$ &   $0.455^{+0.091}_{-0.128}$ &                $6.48^{+4.20}_{-1.95}$ &           $241^{+65}_{-68}$ &   $0.029^{+0.017}_{-0.012}$ &     col \\
      201426077  &        $5658^{+169}_{-135}$ &   $3.888^{+0.096}_{-0.040}$ &  $-0.251^{+0.150}_{-0.150}$ &      $1.91^{+0.15}_{-0.22}$ &   $1.072^{+0.068}_{-0.108}$ &             $0.149^{+0.054}_{-0.024}$ &         $1034^{+86}_{-125}$ &   $0.028^{+0.013}_{-0.008}$ &     lam \\
      201426087  &         $4977^{+100}_{-80}$ &   $2.774^{+0.600}_{-0.600}$ &  $-0.735^{+0.300}_{-0.420}$ &      $5.71^{+6.01}_{-2.63}$ &   $0.915^{+0.116}_{-0.041}$ &          $0.0038^{+0.0165}_{-0.0033}$ &       $1736^{+1709}_{-697}$ &   $0.029^{+0.016}_{-0.010}$ &     rpm \\
      201426089  &        $4229^{+153}_{-153}$ &   $4.751^{+0.055}_{-0.055}$ &  $-0.103^{+0.160}_{-0.240}$ &   $0.523^{+0.065}_{-0.058}$ &   $0.581^{+0.079}_{-0.066}$ &                $3.82^{+1.04}_{-0.82}$ &           $510^{+92}_{-77}$ &   $0.033^{+0.016}_{-0.013}$ &     rpm \\
      201426095  &        $6346^{+154}_{-192}$ &   $4.226^{+0.085}_{-0.170}$ &  $-0.146^{+0.150}_{-0.270}$ &      $1.35^{+0.38}_{-0.16}$ &      $1.17^{+0.13}_{-0.15}$ &                $0.45^{+0.16}_{-0.22}$ &          $504^{+148}_{-66}$ &   $0.028^{+0.016}_{-0.009}$ &     rpm \\
      201426122  &        $3712^{+263}_{-219}$ &   $4.968^{+0.100}_{-0.100}$ &  $-0.103^{+0.200}_{-0.280}$ &   $0.290^{+0.097}_{-0.085}$ &      $0.31^{+0.14}_{-0.12}$ &               $11.03^{+9.79}_{-4.53}$ &          $231^{+104}_{-83}$ &   $0.027^{+0.016}_{-0.010}$ &     col \\
      201426233  &        $5859^{+114}_{-183}$ &   $4.330^{+0.130}_{-0.260}$ &  $-0.133^{+0.200}_{-0.240}$ &      $1.11^{+0.45}_{-0.20}$ &      $0.99^{+0.13}_{-0.10}$ &                $0.67^{+0.50}_{-0.43}$ &         $643^{+268}_{-133}$ &   $0.033^{+0.014}_{-0.014}$ &     rpm \\
      201426240  &        $5657^{+111}_{-111}$ &   $4.467^{+0.072}_{-0.384}$ &  $-0.324^{+0.250}_{-0.350}$ &      $0.90^{+0.39}_{-0.11}$ &   $0.888^{+0.089}_{-0.089}$ &                $1.19^{+0.36}_{-0.73}$ &          $244^{+134}_{-35}$ &   $0.029^{+0.016}_{-0.010}$ &     rpm \\
      201426279  &        $4831^{+117}_{-176}$ &   $4.664^{+0.054}_{-0.030}$ &  $-0.503^{+0.250}_{-0.400}$ &   $0.620^{+0.049}_{-0.063}$ &   $0.659^{+0.059}_{-0.059}$ &                $2.66^{+0.75}_{-0.41}$ &           $443^{+46}_{-48}$ &   $0.024^{+0.025}_{-0.005}$ &     rpm \\
      201426280  &          $4082^{+48}_{-48}$ &   $4.799^{+0.042}_{-0.035}$ &  $-0.045^{+0.150}_{-0.150}$ &   $0.483^{+0.036}_{-0.046}$ &   $0.552^{+0.039}_{-0.055}$ &                $4.64^{+0.91}_{-0.64}$ &           $517^{+51}_{-55}$ &   $0.036^{+0.015}_{-0.015}$ &     rpm \\
      201426327  &          $3786^{+59}_{-59}$ &   $4.954^{+0.063}_{-0.045}$ &  $-0.191^{+0.100}_{-0.100}$ &   $0.334^{+0.035}_{-0.069}$ &   $0.365^{+0.043}_{-0.094}$ &                $9.70^{+4.32}_{-1.92}$ &           $259^{+34}_{-57}$ &   $0.044^{+0.013}_{-0.021}$ &     rpm \\
      201426389  &         $4199^{+124}_{-99}$ &   $4.761^{+0.049}_{-0.035}$ &  $-0.073^{+0.160}_{-0.240}$ &   $0.523^{+0.043}_{-0.054}$ &   $0.573^{+0.048}_{-0.048}$ &                $4.01^{+1.01}_{-0.60}$ &           $398^{+39}_{-49}$ &   $0.030^{+0.015}_{-0.011}$ &     rpm \\
      201426415  &        $3967^{+267}_{-191}$ &   $4.849^{+0.075}_{-0.090}$ &  $-0.055^{+0.150}_{-0.240}$ &   $0.428^{+0.091}_{-0.075}$ &   $0.471^{+0.111}_{-0.092}$ &                $5.92^{+2.64}_{-2.11}$ &           $272^{+85}_{-55}$ &   $0.032^{+0.012}_{-0.012}$ &     rpm \\
      201426494  &        $4115^{+237}_{-237}$ &   $4.803^{+0.080}_{-0.080}$ &  $-0.143^{+0.200}_{-0.240}$ &   $0.469^{+0.103}_{-0.085}$ &   $0.536^{+0.088}_{-0.088}$ &                $4.71^{+1.94}_{-1.38}$ &          $453^{+125}_{-98}$ &   $0.032^{+0.017}_{-0.012}$ &     rpm \\
      201426531  &          $3868^{+38}_{-38}$ &   $4.898^{+0.020}_{-0.030}$ &  $-0.140^{+0.100}_{-0.050}$ &   $0.381^{+0.033}_{-0.011}$ &   $0.423^{+0.041}_{-0.014}$ &                $7.61^{+0.46}_{-1.13}$ &           $270^{+24}_{-12}$ &   $0.057^{+0.001}_{-0.018}$ &     rpm \\
      201426541  &          $3551^{+35}_{-42}$ &   $5.021^{+0.036}_{-0.042}$ &   $0.044^{+0.200}_{-0.160}$ &   $0.254^{+0.036}_{-0.020}$ &   $0.248^{+0.046}_{-0.018}$ &               $15.26^{+2.26}_{-3.36}$ &           $144^{+20}_{-11}$ &   $0.023^{+0.015}_{-0.005}$ &     rpm \\
      201426580  &         $5179^{+144}_{-82}$ &   $3.236^{+0.693}_{-0.990}$ &  $-0.949^{+0.400}_{-1.440}$ &      $3.21^{+6.98}_{-2.09}$ &   $0.876^{+0.026}_{-0.077}$ &             $0.018^{+0.172}_{-0.018}$ &       $2156^{+4483}_{-988}$ &   $0.030^{+0.014}_{-0.012}$ &     rpm \\
      201426597  &          $3872^{+46}_{-46}$ &   $4.893^{+0.049}_{-0.035}$ &  $-0.113^{+0.120}_{-0.120}$ &   $0.383^{+0.038}_{-0.047}$ &   $0.433^{+0.046}_{-0.056}$ &                $7.33^{+2.01}_{-1.16}$ &           $333^{+38}_{-47}$ &   $0.035^{+0.013}_{-0.013}$ &     rpm \\
      201426615  &        $3748^{+386}_{-322}$ &   $4.959^{+0.156}_{-0.156}$ &  $-0.133^{+0.200}_{-0.280}$ &   $0.283^{+0.147}_{-0.097}$ &      $0.32^{+0.18}_{-0.15}$ &              $10.82^{+12.31}_{-5.76}$ &         $409^{+288}_{-193}$ &   $0.029^{+0.017}_{-0.012}$ &     col \\
          \ldots & \ldots & \ldots & \ldots & \ldots & \ldots & \ldots & \ldots & \ldots & \ldots \\

%% file: epic_arxiv_v4.bbl
\newcommand{\SortNoop}[1]{}
\begin{thebibliography}{}
\expandafter\ifx\csname natexlab\endcsname\relax\def\natexlab#1{#1}\fi

\bibitem[{{Ahn} {et~al.}(2012){Ahn}, {Alexandroff}, {Allende Prieto},
  {Anderson}, {Anderton}, {Andrews}, {Aubourg}, {Bailey}, {Balbinot}, {Barnes},
  \& et~al.}]{ahn12}
{Ahn}, C.~P., {Alexandroff}, R., {Allende Prieto}, C., {et~al.} 2012, \apjs,
  203, 21

\bibitem[{{Aigrain} {et~al.}(2015){Aigrain}, {Hodgkin}, {Irwin}, {Lewis}, \&
  {Roberts}}]{aigrain15}
{Aigrain}, S., {Hodgkin}, S.~T., {Irwin}, M.~J., {Lewis}, J.~R., \& {Roberts},
  S.~J. 2015, \mnras, 447, 2880

\bibitem[{{Alam} {et~al.}(2015){Alam}, {Albareti}, {Allende Prieto}, {Anders},
  {Anderson}, {Anderton}, {Andrews}, {Armengaud}, {Aubourg}, {Bailey}, \&
  et~al.}]{alam15}
{Alam}, S., {Albareti}, F.~D., {Allende Prieto}, C., {et~al.} 2015, \apjs, 219,
  12

\bibitem[{{Allende Prieto} {et~al.}(2008){Allende Prieto}, {Majewski},
  {Schiavon}, {Cunha}, {Frinchaboy}, {Holtzman}, {Johnston}, {Shetrone},
  {Skrutskie}, {Smith}, \& {Wilson}}]{allende08}
{Allende Prieto}, C., {Majewski}, S.~R., {Schiavon}, R., {et~al.} 2008,
  Astronomische Nachrichten, 329, 1018

\bibitem[{{Angus} {et~al.}(2016){Angus}, {Foreman-Mackey}, \&
  {Johnson}}]{angus15}
{Angus}, R., {Foreman-Mackey}, D., \& {Johnson}, J.~A. 2016, \apj, 818, 109

\bibitem[{{Armstrong} {et~al.}(2015){Armstrong}, {Kirk}, {Lam}, {McCormac},
  {Walker}, {Brown}, {Osborn}, {Pollacco}, \& {Spake}}]{armstrong15}
{Armstrong}, D.~J., {Kirk}, J., {Lam}, K.~W.~F., {et~al.} 2015, \aap, 579, A19

\bibitem[{{Batalha} {et~al.}(2010){Batalha}, {Borucki}, {Koch}, {Bryson},
  {Haas}, {Brown}, {Caldwell}, {Hall}, {Gilliland}, {Latham}, {Meibom}, \&
  {Monet}}]{batalha10}
{Batalha}, N.~M., {Borucki}, W.~J., {Koch}, D.~G., {et~al.} 2010, \apjl, 713,
  L109

\bibitem[{{Batalha} {et~al.}(2013){Batalha}, {Rowe}, {Bryson}, {Barclay},
  {Burke}, {Caldwell}, {Christiansen}, {Mullally}, {Thompson}, {Brown},
  {Dupree}, {Fabrycky}, {Ford}, {Fortney}, {Gilliland}, {Isaacson}, {Latham},
  {Marcy}, {Quinn}, {Ragozzine}, {Shporer}, {Borucki}, {Ciardi}, {Gautier},
  {Haas}, {Jenkins}, {Koch}, {Lissauer}, {Rapin}, {Basri}, {Boss}, {Buchhave},
  {Carter}, {Charbonneau}, {Christensen-Dalsgaard}, {Clarke}, {Cochran},
  {Demory}, {Desert}, {Devore}, {Doyle}, {Esquerdo}, {Everett}, {Fressin},
  {Geary}, {Girouard}, {Gould}, {Hall}, {Holman}, {Howard}, {Howell},
  {Ibrahim}, {Kinemuchi}, {Kjeldsen}, {Klaus}, {Li}, {Lucas}, {Meibom},
  {Morris}, {Pr{\v s}a}, {Quintana}, {Sanderfer}, {Sasselov}, {Seader},
  {Smith}, {Steffen}, {Still}, {Stumpe}, {Tarter}, {Tenenbaum}, {Torres},
  {Twicken}, {Uddin}, {Van Cleve}, {Walkowicz}, \& {Welsh}}]{batalha12}
{Batalha}, N.~M., {Rowe}, J.~F., {Bryson}, S.~T., {et~al.} 2013, \apjs, 204, 24

\bibitem[{{Bessell}(2000)}]{bessell00}
{Bessell}, M.~S. 2000, \pasp, 112, 961

\bibitem[{{Bilir} {et~al.}(2008){Bilir}, {Ak}, {Karaali}, {Cabrera-Lavers},
  {Chonis}, \& {Gaskell}}]{bilir08}
{Bilir}, S., {Ak}, S., {Karaali}, S., {et~al.} 2008, \mnras, 384, 1178

\bibitem[{{Bilir} {et~al.}(2005){Bilir}, {Karaali}, \& {Tun{\c c}el}}]{bilir05}
{Bilir}, S., {Karaali}, S., \& {Tun{\c c}el}, S. 2005, Astronomische
  Nachrichten, 326, 321

\bibitem[{{Borucki} {et~al.}(2010){Borucki}, {Koch}, {Basri}, {Batalha},
  {Brown}, {Caldwell}, {Caldwell}, {Christensen-Dalsgaard}, {Cochran},
  {DeVore}, {Dunham}, {Dupree}, {Gautier}, {Geary}, {Gilliland}, {Gould},
  {Howell}, {Jenkins}, {Kondo}, {Latham}, {Marcy}, {Meibom}, {Kjeldsen},
  {Lissauer}, {Monet}, {Morrison}, {Sasselov}, {Tarter}, {Boss}, {Brownlee},
  {Owen}, {Buzasi}, {Charbonneau}, {Doyle}, {Fortney}, {Ford}, {Holman},
  {Seager}, {Steffen}, {Welsh}, {Rowe}, {Anderson}, {Buchhave}, {Ciardi},
  {Walkowicz}, {Sherry}, {Horch}, {Isaacson}, {Everett}, {Fischer}, {Torres},
  {Johnson}, {Endl}, {MacQueen}, {Bryson}, {Dotson}, {Haas}, {Kolodziejczak},
  {Van Cleve}, {Chandrasekaran}, {Twicken}, {Quintana}, {Clarke}, {Allen},
  {Li}, {Wu}, {Tenenbaum}, {Verner}, {Bruhweiler}, {Barnes}, \&
  {Prsa}}]{borucki10}
{Borucki}, W.~J., {Koch}, D., {Basri}, G., {et~al.} 2010, Science, 327, 977

\bibitem[{{Borucki} {et~al.}(2011{\natexlab{a}}){Borucki}, {Koch}, {Basri},
  {Batalha}, {Boss}, {Brown}, {Caldwell}, {Christensen-Dalsgaard}, {Cochran},
  {DeVore}, {Dunham}, {Dupree}, {Gautier}, {Geary}, {Gilliland}, {Gould},
  {Howell}, {Jenkins}, {Kjeldsen}, {Latham}, {Lissauer}, {Marcy}, {Monet},
  {Sasselov}, {Tarter}, {Charbonneau}, {Doyle}, {Ford}, {Fortney}, {Holman},
  {Seager}, {Steffen}, {Welsh}, {Allen}, {Bryson}, {Buchhave},
  {Chandrasekaran}, {Christiansen}, {Ciardi}, {Clarke}, {Dotson}, {Endl},
  {Fischer}, {Fressin}, {Haas}, {Horch}, {Howard}, {Isaacson}, {Kolodziejczak},
  {Li}, {MacQueen}, {Meibom}, {Prsa}, {Quintana}, {Rowe}, {Sherry},
  {Tenenbaum}, {Torres}, {Twicken}, {Van Cleve}, {Walkowicz}, \&
  {Wu}}]{borucki11b}
{Borucki}, W.~J., {Koch}, D.~G., {Basri}, G., {et~al.} 2011{\natexlab{a}},
  \apj, 728, 117

\bibitem[{{Borucki} {et~al.}(2011{\natexlab{b}}){Borucki}, {Koch}, {Basri},
  {Batalha}, {Brown}, {Bryson}, {Caldwell}, {Christensen-Dalsgaard}, {Cochran},
  {DeVore}, {Dunham}, {Gautier}, {Geary}, {Gilliland}, {Gould}, {Howell},
  {Jenkins}, {Latham}, {Lissauer}, {Marcy}, {Rowe}, {Sasselov}, {Boss},
  {Charbonneau}, {Ciardi}, {Doyle}, {Dupree}, {Ford}, {Fortney}, {Holman},
  {Seager}, {Steffen}, {Tarter}, {Welsh}, {Allen}, {Buchhave}, {Christiansen},
  {Clarke}, {Das}, {D{\'e}sert}, {Endl}, {Fabrycky}, {Fressin}, {Haas},
  {Horch}, {Howard}, {Isaacson}, {Kjeldsen}, {Kolodziejczak}, {Kulesa}, {Li},
  {Lucas}, {Machalek}, {McCarthy}, {MacQueen}, {Meibom}, {Miquel}, {Prsa},
  {Quinn}, {Quintana}, {Ragozzine}, {Sherry}, {Shporer}, {Tenenbaum}, {Torres},
  {Twicken}, {Van Cleve}, {Walkowicz}, {Witteborn}, \& {Still}}]{borucki11}
---. 2011{\natexlab{b}}, \apj, 736, 19

\bibitem[{{Boyajian} {et~al.}(2012{\natexlab{a}}){Boyajian}, {McAlister}, {van
  Belle}, {Gies}, {ten Brummelaar}, {von Braun}, {Farrington}, {Goldfinger},
  {O'Brien}, {Parks}, {Richardson}, {Ridgway}, {Schaefer}, {Sturmann},
  {Sturmann}, {Touhami}, {Turner}, \& {White}}]{boyajian12}
{Boyajian}, T.~S., {McAlister}, H.~A., {van Belle}, G., {et~al.}
  2012{\natexlab{a}}, \apj, 746, 101

\bibitem[{{Boyajian} {et~al.}(2012{\natexlab{b}}){Boyajian}, {von Braun}, {van
  Belle}, {McAlister}, {ten Brummelaar}, {Kane}, {Muirhead}, {Jones}, {White},
  {Schaefer}, {Ciardi}, {Henry}, {L{\'o}pez-Morales}, {Ridgway}, {Gies}, {Jao},
  {Rojas-Ayala}, {Parks}, {Sturmann}, {Sturmann}, {Turner}, {Farrington},
  {Goldfinger}, \& {Berger}}]{boyajian12b}
{Boyajian}, T.~S., {von Braun}, K., {van Belle}, G., {et~al.}
  2012{\natexlab{b}}, \apj, 757, 112

\bibitem[{{Boyajian} {et~al.}(2013){Boyajian}, {von Braun}, {van Belle},
  {Farrington}, {Schaefer}, {Jones}, {White}, {McAlister}, {ten Brummelaar},
  {Ridgway}, {Gies}, {Sturmann}, {Sturmann}, {Turner}, {Goldfinger}, \&
  {Vargas}}]{boyajian13}
---. 2013, \apj, 771, 40

\bibitem[{{Bressan} {et~al.}(2012){Bressan}, {Marigo}, {Girardi}, {Salasnich},
  {Dal Cero}, {Rubele}, \& {Nanni}}]{bressan12}
{Bressan}, A., {Marigo}, P., {Girardi}, L., {et~al.} 2012, \mnras, 427, 127

\bibitem[{{Brown} {et~al.}(2011){Brown}, {Latham}, {Everett}, \&
  {Esquerdo}}]{brown11}
{Brown}, T.~M., {Latham}, D.~W., {Everett}, M.~E., \& {Esquerdo}, G.~A. 2011,
  \aj, 142, 112

\bibitem[{{Bruntt} {et~al.}(2010){Bruntt}, {Bedding}, {Quirion}, {Lo Curto},
  {Carrier}, {Smalley}, {Dall}, {Arentoft}, {Bazot}, \& {Butler}}]{bruntt10}
{Bruntt}, H., {Bedding}, T.~R., {Quirion}, P.-O., {et~al.} 2010, \mnras, 405,
  1907

\bibitem[{{Buchhave} {et~al.}(2012){Buchhave}, {Latham}, {Johansen},
  {Bizzarro}, {Torres}, {Rowe}, {Batalha}, {Borucki}, {Brugamyer}, {Caldwell},
  {Bryson}, {Ciardi}, {Cochran}, {Endl}, {Esquerdo}, {Ford}, {Geary},
  {Gilliland}, {Hansen}, {Isaacson}, {Laird}, {Lucas}, {Marcy}, {Morse},
  {Robertson}, {Shporer}, {Stefanik}, {Still}, \& {Quinn}}]{buchhave12}
{Buchhave}, L.~A., {Latham}, D.~W., {Johansen}, A., {et~al.} 2012, \nat, 486,
  375

\bibitem[{{Buchhave} {et~al.}(2014){Buchhave}, {Bizzarro}, {Latham},
  {Sasselov}, {Cochran}, {Endl}, {Isaacson}, {Juncher}, \&
  {Marcy}}]{buchhave14}
{Buchhave}, L.~A., {Bizzarro}, M., {Latham}, D.~W., {et~al.} 2014, \nat, 509,
  593

\bibitem[{{Buysschaert} {et~al.}(2015){Buysschaert}, {Aerts}, {Bloemen},
  {Debosscher}, {Neiner}, {Briquet}, {Vos}, {P{\'a}pics}, {Manick}, {Schmid},
  {Van Winckel}, \& {Tkachenko}}]{buysschaert15}
{Buysschaert}, B., {Aerts}, C., {Bloemen}, S., {et~al.} 2015, \mnras, 453, 89

\bibitem[{{Carter} {et~al.}(2011){Carter}, {Fabrycky}, {Ragozzine}, {Holman},
  {Quinn}, {Latham}, {Buchhave}, {Van Cleve}, {Cochran}, {Cote}, {Endl},
  {Ford}, {Haas}, {Jenkins}, {Koch}, {Li}, {Lissauer}, {MacQueen}, {Middour},
  {Orosz}, {Rowe}, {Steffen}, \& {Welsh}}]{carter11}
{Carter}, J.~A., {Fabrycky}, D.~C., {Ragozzine}, D., {et~al.} 2011, Science,
  331, 562

\bibitem[{{Casagrande} {et~al.}(2011){Casagrande}, {Sch{\"o}nrich}, {Asplund},
  {Cassisi}, {Ram{\'{\i}}rez}, {Mel{\'e}ndez}, {Bensby}, \&
  {Feltzing}}]{casagrande11}
{Casagrande}, L., {Sch{\"o}nrich}, R., {Asplund}, M., {et~al.} 2011, \aap, 530,
  A138

\bibitem[{{Casagrande} {et~al.}(2014){Casagrande}, {Silva Aguirre}, {Stello},
  {Huber}, {Serenelli}, {Cassisi}, {Dotter}, {Milone}, {Hodgkin}, {Marino},
  {Lund}, {Pietrinferni}, {Asplund}, {Feltzing}, {Flynn}, {Grundahl}, {Nissen},
  {Sch{\"o}nrich}, {Schlesinger}, \& {Wang}}]{casagrande14}
{Casagrande}, L., {Silva Aguirre}, V., {Stello}, D., {et~al.} 2014, \apj, 787,
  110

\bibitem[{{Chaplin} {et~al.}(2011){Chaplin}, {Kjeldsen},
  {Christensen-Dalsgaard}, {Basu}, {Miglio}, {Appourchaux}, {Bedding},
  {Elsworth}, {Garc{\'{\i}}a}, {Gilliland}, {Girardi}, {Houdek}, {Karoff},
  {Kawaler}, {Metcalfe}, {Molenda-{\.Z}akowicz}, {Monteiro}, {Thompson},
  {Verner}, {Ballot}, {Bonanno}, {Brand{\~a}o}, {Broomhall}, {Bruntt},
  {Campante}, {Corsaro}, {Creevey}, {Do{\u g}an}, {Esch}, {Gai}, {Gaulme},
  {Hale}, {Handberg}, {Hekker}, {Huber}, {Jim{\'e}nez}, {Mathur}, {Mazumdar},
  {Mosser}, {New}, {Pinsonneault}, {Pricopi}, {Quirion}, {R{\'e}gulo},
  {Salabert}, {Serenelli}, {Silva Aguirre}, {Sousa}, {Stello}, {Stevens},
  {Suran}, {Uytterhoeven}, {White}, {Borucki}, {Brown}, {Jenkins}, {Kinemuchi},
  {Van Cleve}, \& {Klaus}}]{chaplin11a}
{Chaplin}, W.~J., {Kjeldsen}, H., {Christensen-Dalsgaard}, J., {et~al.} 2011,
  Science, 332, 213

\bibitem[{{Chaplin} {et~al.}(2015){Chaplin}, {Lund}, {Handberg}, {Basu},
  {Buchhave}, {Campante}, {Davies}, {Huber}, {Latham}, {Latham}, {Serenelli},
  {Antia}, {Appourchaux}, {Ball}, {Benomar}, {Casagrande},
  {Christensen-Dalsgaard}, {Coelho}, {Creevey}, {Elsworth}, {Garc{\'{\i}}a},
  {Gaulme}, {Hekker}, {Kallinger}, {Karoff}, {Kawaler}, {Kjeldsen},
  {Lundkvist}, {Marcadon}, {Mathur}, {Miglio}, {Mosser}, {R{\'e}gulo},
  {Roxburgh}, {Silva Aguirre}, {Stello}, {Verma}, {White}, {Bedding},
  {Barclay}, {Buzasi}, {Dehuevels}, {Gizon}, {Houdek}, {Howell}, {Salabert}, \&
  {Soderblom}}]{chaplin15}
{Chaplin}, W.~J., {Lund}, M.~N., {Handberg}, R., {et~al.} 2015, \pasp, 127,
  1038

\bibitem[{{Crossfield} {et~al.}(2015){Crossfield}, {Petigura}, {Schlieder},
  {Howard}, {Fulton}, {Aller}, {Ciardi}, {L{\'e}pine}, {Barclay}, {de Pater},
  {de Kleer}, {Quintana}, {Christiansen}, {Schlafly}, {Kaltenegger}, {Crepp},
  {Henning}, {Obermeier}, {Deacon}, {Weiss}, {Isaacson}, {Hansen}, {Liu},
  {Greene}, {Howell}, {Barman}, \& {Mordasini}}]{crossfield15}
{Crossfield}, I.~J.~M., {Petigura}, E., {Schlieder}, J.~E., {et~al.} 2015,
  \apj, 804, 10

\bibitem[{{De Silva} {et~al.}(2015){De Silva}, {Freeman}, {Bland-Hawthorn},
  {Martell}, {de Boer}, {Asplund}, {Keller}, {Sharma}, {Zucker}, {Zwitter},
  {Anguiano}, {Bacigalupo}, {Bayliss}, {Beavis}, {Bergemann}, {Campbell},
  {Cannon}, {Carollo}, {Casagrande}, {Casey}, {Da Costa}, {D'Orazi}, {Dotter},
  {Duong}, {Heger}, {Ireland}, {Kafle}, {Kos}, {Lattanzio}, {Lewis}, {Lin},
  {Lind}, {Munari}, {Nataf}, {O'Toole}, {Parker}, {Reid}, {Schlesinger},
  {Sheinis}, {Simpson}, {Stello}, {Ting}, {Traven}, {Watson}, {Wittenmyer},
  {Yong}, \& {{\v Z}erjal}}]{desilva15}
{De Silva}, G.~M., {Freeman}, K.~C., {Bland-Hawthorn}, J., {et~al.} 2015,
  \mnras, 449, 2604

\bibitem[{{Dotter} {et~al.}(2008){Dotter}, {Chaboyer}, {Jevremovi{\'c}},
  {Kostov}, {Baron}, \& {Ferguson}}]{dotter08}
{Dotter}, A., {Chaboyer}, B., {Jevremovi{\'c}}, D., {et~al.} 2008, \apjs, 178,
  89

\bibitem[{{Foreman-Mackey} {et~al.}(2015){Foreman-Mackey}, {Montet}, {Hogg},
  {Morton}, {Wang}, \& {Sch{\"o}lkopf}}]{dfm15}
{Foreman-Mackey}, D., {Montet}, B.~T., {Hogg}, D.~W., {et~al.} 2015, \apj, 806,
  215

\bibitem[{{Fressin} {et~al.}(2013){Fressin}, {Torres}, {Charbonneau}, {Bryson},
  {Christiansen}, {Dressing}, {Jenkins}, {Walkowicz}, \& {Batalha}}]{fressin13}
{Fressin}, F., {Torres}, G., {Charbonneau}, D., {et~al.} 2013, \apj, 766, 81

\bibitem[{{Girardi} {et~al.}(2000){Girardi}, {Bressan}, {Bertelli}, \&
  {Chiosi}}]{girardi00}
{Girardi}, L., {Bressan}, A., {Bertelli}, G., \& {Chiosi}, C. 2000, \aaps, 141,
  371

\bibitem[{{Gould} \& {Morgan}(2003)}]{gould03}
{Gould}, A., \& {Morgan}, C.~W. 2003, \apj, 585, 1056

\bibitem[{{Henderson} {et~al.}(2015){Henderson}, {Penny}, {Street}, {Bennett},
  {Hogg}, {Poleski}, {Barclay}, {Barentsen}, {Howell}, {Udalski},
  {Szyma{\'n}ski}, {Skowron}, {Mr{\'o}z}, {Koz{\l}owski}, {Wyrzykowski},
  {Pietrukowicz}, {Soszy{\'n}ski}, {Ulaczyk}, {Pawlak}, {Sumi}, {Abe},
  {Asakura}, {Barry}, {Bhattacharya}, {Bond}, {Donachie}, {Freeman}, {Fukui},
  {Hirao}, {Itow}, {Koshimoto}, {Li}, {Ling}, {Masuda}, {Matsubara}, {Muraki},
  {Nagakane}, {Ohnishi}, {Oyokawa}, {Rattenbury}, {Saito}, {Sharan},
  {Sullivan}, {Tristram}, {Yonehara}, {Bachelet}, {Bramich}, {Cassan},
  {Dominik}, {Figuera Jaimes}, {Horne}, {Hundertmark}, {Mao}, {Ranc},
  {Schmidt}, {Snodgrass}, {Steele}, {Tsapras}, {Wambsganss}, {Akeson},
  {Batista}, {Beaulieu}, {Beichman}, {Bozza}, {Bryden}, {Ciardi}, {Cole},
  {Coutures}, {Dong}, {Foreman-Mackey}, {Fouqu{\'e}}, {Gaudi}, {Kerins},
  {Korhonen}, {J{\o}rgensen}, {Lang}, {Lineweaver}, {Marquette}, {Mogavero},
  {Morales}, {Nataf}, {Pogge}, {Santerne}, {Shvartzvald}, {Suzuki}, {Tamura},
  {Tisserand}, {Wang}, \& {Zhu}}]{henderson15}
{Henderson}, C.~B., {Penny}, M., {Street}, R.~A., {et~al.} 2015, ArXiv
  e-prints, arXiv:1512.09142

\bibitem[{{Hermes} {et~al.}(2014){Hermes}, {Charpinet}, {Barclay}, {Pak{\v
  s}tien{\.e}}, {Mullally}, {Kawaler}, {Bloemen}, {Castanheira}, {Winget},
  {Montgomery}, {Van Grootel}, {Huber}, {Still}, {Howell}, {Caldwell}, {Haas},
  \& {Bryson}}]{hermes14}
{Hermes}, J.~J., {Charpinet}, S., {Barclay}, T., {et~al.} 2014, \apj, 789, 85

\bibitem[{{H{\o}g} {et~al.}(2000){H{\o}g}, {Fabricius}, {Makarov}, {Urban},
  {Corbin}, {Wycoff}, {Bastian}, {Schwekendiek}, \& {Wicenec}}]{hog00}
{H{\o}g}, E., {Fabricius}, C., {Makarov}, V.~V., {et~al.} 2000, \aap, 355, L27

\bibitem[{{Howard} {et~al.}(2012){Howard}, {Marcy}, {Bryson}, {Jenkins},
  {Rowe}, {Batalha}, {Borucki}, {Koch}, {Dunham}, {Gautier}, {Van Cleve},
  {Cochran}, {Latham}, {Lissauer}, {Torres}, {Brown}, {Gilliland}, {Buchhave},
  {Caldwell}, {Christensen-Dalsgaard}, {Ciardi}, {Fressin}, {Haas}, {Howell},
  {Kjeldsen}, {Seager}, {Rogers}, {Sasselov}, {Steffen}, {Basri},
  {Charbonneau}, {Christiansen}, {Clarke}, {Dupree}, {Fabrycky}, {Fischer},
  {Ford}, {Fortney}, {Tarter}, {Girouard}, {Holman}, {Johnson}, {Klaus},
  {Machalek}, {Moorhead}, {Morehead}, {Ragozzine}, {Tenenbaum}, {Twicken},
  {Quinn}, {Isaacson}, {Shporer}, {Lucas}, {Walkowicz}, {Welsh}, {Boss},
  {Devore}, {Gould}, {Smith}, {Morris}, {Prsa}, {Morton}, {Still}, {Thompson},
  {Mullally}, {Endl}, \& {MacQueen}}]{howard11}
{Howard}, A.~W., {Marcy}, G.~W., {Bryson}, S.~T., {et~al.} 2012, \apjs, 201, 15

\bibitem[{{Howell} {et~al.}(2012){Howell}, {Rowe}, {Bryson}, {Quinn}, {Marcy},
  {Isaacson}, {Ciardi}, {Chaplin}, {Metcalfe}, {Monteiro}, {Appourchaux},
  {Basu}, {Creevey}, {Gilliland}, {Quirion}, {Stello}, {Kjeldsen},
  {Christensen-Dalsgaard}, {Elsworth}, {Garc{\'{\i}}a}, {Houdek}, {Karoff},
  {Molenda-{\.Z}akowicz}, {Thompson}, {Verner}, {Torres}, {Fressin}, {Crepp},
  {Adams}, {Dupree}, {Sasselov}, {Dressing}, {Borucki}, {Koch}, {Lissauer},
  {Latham}, {Buchhave}, {Gautier}, {Everett}, {Horch}, {Batalha}, {Dunham},
  {Szkody}, {Silva}, {Mighell}, {Holberg}, {Ballot}, {Bedding}, {Bruntt},
  {Campante}, {Handberg}, {Hekker}, {Huber}, {Mathur}, {Mosser}, {R{\'e}gulo},
  {White}, {Christiansen}, {Middour}, {Haas}, {Hall}, {Jenkins}, {McCaulif},
  {Fanelli}, {Kulesa}, {McCarthy}, \& {Henze}}]{howell12}
{Howell}, S.~B., {Rowe}, J.~F., {Bryson}, S.~T., {et~al.} 2012, \apj, 746, 123

\bibitem[{{Howell} {et~al.}(2014){Howell}, {Sobeck}, {Haas}, {Still},
  {Barclay}, {Mullally}, {Troeltzsch}, {Aigrain}, {Bryson}, {Caldwell},
  {Chaplin}, {Cochran}, {Huber}, {Marcy}, {Miglio}, {Najita}, {Smith},
  {Twicken}, \& {Fortney}}]{howell14}
{Howell}, S.~B., {Sobeck}, C., {Haas}, M., {et~al.} 2014, \pasp, 126, 398

\bibitem[{{Huang} {et~al.}(2015){Huang}, {Penev}, {Hartman}, {Bakos}, {Bhatti},
  {Domsa}, \& {de Val-Borro}}]{huang15}
{Huang}, C.~X., {Penev}, K., {Hartman}, J.~D., {et~al.} 2015, \mnras, 454, 4159

\bibitem[{{Huber} {et~al.}(2009){Huber}, {Stello}, {Bedding}, {Chaplin},
  {Arentoft}, {Quirion}, \& {Kjeldsen}}]{huber09}
{Huber}, D., {Stello}, D., {Bedding}, T.~R., {et~al.} 2009, Communications in
  Asteroseismology, 160, 74

\bibitem[{{Huber} {et~al.}(2013){Huber}, {Chaplin}, {Christensen-Dalsgaard},
  {Gilliland}, {Kjeldsen}, {Buchhave}, {Fischer}, {Lissauer}, {Rowe},
  {Sanchis-Ojeda}, {Basu}, {Handberg}, {Hekker}, {Howard}, {Isaacson},
  {Karoff}, {Latham}, {Lund}, {Lundkvist}, {Marcy}, {Miglio}, {Silva Aguirre},
  {Stello}, {Arentoft}, {Barclay}, {Bedding}, {Burke}, {Christiansen},
  {Elsworth}, {Haas}, {Kawaler}, {Metcalfe}, {Mullally}, \&
  {Thompson}}]{huber13}
{Huber}, D., {Chaplin}, W.~J., {Christensen-Dalsgaard}, J., {et~al.} 2013,
  \apj, 767, 127

\bibitem[{{Huber} {et~al.}(2014){Huber}, {Silva Aguirre}, {Matthews},
  {Pinsonneault}, {Gaidos}, {Garc{\'{\i}}a}, {Hekker}, {Mathur}, {Mosser},
  {Torres}, {Bastien}, {Basu}, {Bedding}, {Chaplin}, {Demory}, {Fleming},
  {Guo}, {Mann}, {Rowe}, {Serenelli}, {Smith}, \& {Stello}}]{huber14}
{Huber}, D., {Silva Aguirre}, V., {Matthews}, J.~M., {et~al.} 2014, \apjs, 211,
  2

\bibitem[{{Jeffery} \& {Ramsay}(2014)}]{jeffery14}
{Jeffery}, C.~S., \& {Ramsay}, G. 2014, \mnras, 442, L61

\bibitem[{{Johnson} \& {Soderblom}(1987)}]{johnson87}
{Johnson}, D.~R.~H., \& {Soderblom}, D.~R. 1987, \aj, 93, 864

\bibitem[{{Kallinger} {et~al.}(2010){Kallinger}, {Mosser}, {Hekker}, {Huber},
  {Stello}, {Mathur}, {Basu}, {Bedding}, {Chaplin}, {De Ridder}, {Elsworth},
  {Frandsen}, {Garc{\'{\i}}a}, {Gruberbauer}, {Matthews}, {Borucki}, {Bruntt},
  {Christensen-Dalsgaard}, {Gilliland}, {Kjeldsen}, \& {Koch}}]{kallinger10}
{Kallinger}, T., {Mosser}, B., {Hekker}, S., {et~al.} 2010, \aap, 522, A1

\bibitem[{{Koch} {et~al.}(2010){Koch}, {Borucki}, {Basri}, {Batalha}, {Brown},
  {Caldwell}, {Christensen-Dalsgaard}, {Cochran}, {DeVore}, {Dunham},
  {Gautier}, {Geary}, {Gilliland}, {Gould}, {Jenkins}, {Kondo}, {Latham},
  {Lissauer}, {Marcy}, {Monet}, {Sasselov}, {Boss}, {Brownlee}, {Caldwell},
  {Dupree}, {Howell}, {Kjeldsen}, {Meibom}, {Morrison}, {Owen}, {Reitsema},
  {Tarter}, {Bryson}, {Dotson}, {Gazis}, {Haas}, {Kolodziejczak}, {Rowe}, {Van
  Cleve}, {Allen}, {Chandrasekaran}, {Clarke}, {Li}, {Quintana}, {Tenenbaum},
  {Twicken}, \& {Wu}}]{koch10b}
{Koch}, D.~G., {Borucki}, W.~J., {Basri}, G., {et~al.} 2010, \apjl, 713, L79

\bibitem[{{Kordopatis} {et~al.}(2013){Kordopatis}, {Gilmore}, {Steinmetz},
  {Boeche}, {Seabroke}, {Siebert}, {Zwitter}, {Binney}, {de Laverny},
  {Recio-Blanco}, {Williams}, {Piffl}, {Enke}, {Roeser}, {Bijaoui}, {Wyse},
  {Freeman}, {Munari}, {Carrillo}, {Anguiano}, {Burton}, {Campbell}, {Cass},
  {Fiegert}, {Hartley}, {Parker}, {Reid}, {Ritter}, {Russell}, {Stupar},
  {Watson}, {Bienaym{\'e}}, {Bland-Hawthorn}, {Gerhard}, {Gibson}, {Grebel},
  {Helmi}, {Navarro}, {Conrad}, {Famaey}, {Faure}, {Just}, {Kos}, {Matijevi{\v
  c}}, {McMillan}, {Minchev}, {Scholz}, {Sharma}, {Siviero}, {de Boer}, \& {{\v
  Z}erjal}}]{kordopatis13}
{Kordopatis}, G., {Gilmore}, G., {Steinmetz}, M., {et~al.} 2013, \aj, 146, 134

\bibitem[{{Kraus} {et~al.}(2011){Kraus}, {Tucker}, {Thompson}, {Craine}, \&
  {Hillenbrand}}]{kraus11}
{Kraus}, A.~L., {Tucker}, R.~A., {Thompson}, M.~I., {Craine}, E.~R., \&
  {Hillenbrand}, L.~A. 2011, \apj, 728, 48

\bibitem[{{Kurtz} {et~al.}(2016){Kurtz}, {Bowman}, {Ebo}, {Moskalik},
  {Handberg}, \& {Lund}}]{kurtz16}
{Kurtz}, D.~W., {Bowman}, D.~M., {Ebo}, S.~J., {et~al.} 2016, \mnras, 455, 1237

\bibitem[{{LaCourse} {et~al.}(2015){LaCourse}, {Jek}, {Jacobs}, {Winarski},
  {Boyajian}, {Rappaport}, {Sanchis-Ojeda}, {Conroy}, {Nelson}, {Barclay},
  {Fischer}, {Schmitt}, {Wang}, {Stassun}, {Pepper}, {Coughlin}, {Shporer}, \&
  {Pr{\v s}a}}]{laCourse15}
{LaCourse}, D.~M., {Jek}, K.~J., {Jacobs}, T.~L., {et~al.} 2015, \mnras, 452,
  3561

\bibitem[{{Lund} {et~al.}(2015){Lund}, {Handberg}, {Davies}, {Chaplin}, \&
  {Jones}}]{lund15}
{Lund}, M.~N., {Handberg}, R., {Davies}, G.~R., {Chaplin}, W.~J., \& {Jones},
  C.~D. 2015, \apj, 806, 30

\bibitem[{{Luo} {et~al.}(2015){Luo}, {Zhao}, {Zhao}, {Deng}, {Liu}, {Jing},
  {Wang}, {Zhang}, {Shi}, {Cui}, {Chu}, {Li}, {Bai}, {Wu}, {Cai}, {Cao}, {Cao},
  {Carlin}, {Chen}, {Chen}, {Chen}, {Chen}, {Chen}, {Chen}, {Chen},
  {Christlieb}, {Chu}, {Cui}, {Dong}, {Du}, {Fan}, {Feng}, {Fu}, {Gao}, {Gong},
  {Gu}, {Guo}, {Han}, {He}, {Hou}, {Hou}, {Hou}, {Hu}, {Hu}, {Hu}, {Huo},
  {Jia}, {Jiang}, {Jiang}, {Jiang}, {Jin}, {Kong}, {Kong}, {Lei}, {Li}, {Li},
  {Li}, {Li}, {Li}, {Li}, {Li}, {Li}, {Li}, {Li}, {Li}, {Li}, {Liang}, {Lin},
  {Liu}, {Liu}, {Liu}, {Liu}, {Lu}, {Luo}, {Mao}, {Newberg}, {Ni}, {Qi}, {Qi},
  {Shen}, {Shi}, {Song}, {Song}, {Su}, {Su}, {Tang}, {Tao}, {Tian}, {Wang},
  {Wang}, {Wang}, {Wang}, {Wang}, {Wang}, {Wang}, {Wang}, {Wang}, {Wang},
  {Wang}, {Wang}, {Wang}, {Wang}, {Wang}, {Wang}, {Wang}, {Wang}, {Wang},
  {Wang}, {Wei}, {Wei}, {Wu}, {Wu}, {Wu}, {Wu}, {Xing}, {Xu}, {Xu}, {Xu},
  {Yan}, {Yang}, {Yang}, {Yang}, {Yang}, {Yao}, {Yu}, {Yuan}, {Yuan}, {Yuan},
  {Yuan}, {Zhai}, {Zhang}, {Zhang}, {Zhang}, {Zhang}, {Zhang}, {Zhang},
  {Zhang}, {Zhang}, {Zhao}, {Zhou}, {Zhou}, {Zhu}, {Zhu}, {Zou}, \&
  {Zuo}}]{luo15}
{Luo}, A.-L., {Zhao}, Y.-H., {Zhao}, G., {et~al.} 2015, Research in Astronomy
  and Astrophysics, 15, 1095

\bibitem[{{Mann} {et~al.}(2015){Mann}, {Feiden}, {Gaidos}, {Boyajian}, \& {von
  Braun}}]{mann15}
{Mann}, A.~W., {Feiden}, G.~A., {Gaidos}, E., {Boyajian}, T., \& {von Braun},
  K. 2015, \apj, 804, 64

\bibitem[{{Marigo} \& {Girardi}(2007)}]{marigo07}
{Marigo}, P., \& {Girardi}, L. 2007, \aap, 469, 239

\bibitem[{{Marigo} {et~al.}(2008){Marigo}, {Girardi}, {Bressan}, {Groenewegen},
  {Silva}, \& {Granato}}]{marigo08}
{Marigo}, P., {Girardi}, L., {Bressan}, A., {et~al.} 2008, \aap, 482, 883

\bibitem[{{Miglio} {et~al.}(2013){Miglio}, {Chiappini}, {Morel}, {Barbieri},
  {Chaplin}, {Girardi}, {Montalb{\'a}n}, {Valentini}, {Mosser}, {Baudin},
  {Casagrande}, {Fossati}, {Aguirre}, \& {Baglin}}]{miglio12}
{Miglio}, A., {Chiappini}, C., {Morel}, T., {et~al.} 2013, \mnras, 429, 423

\bibitem[{{Moln{\'a}r} {et~al.}(2015){Moln{\'a}r}, {Szab{\'o}}, {Moskalik},
  {Nemec}, {Guggenberger}, {Smolec}, {Poleski}, {Plachy}, {Kolenberg}, \&
  {Koll{\'a}th}}]{molnar15}
{Moln{\'a}r}, L., {Szab{\'o}}, R., {Moskalik}, P.~A., {et~al.} 2015, \mnras,
  452, 4283

\bibitem[{{Montet} {et~al.}(2015){Montet}, {Morton}, {Foreman-Mackey},
  {Johnson}, {Hogg}, {Bowler}, {Latham}, {Bieryla}, \& {Mann}}]{montet15}
{Montet}, B.~T., {Morton}, T.~D., {Foreman-Mackey}, D., {et~al.} 2015, \apj,
  809, 25

\bibitem[{{Morton}(2015)}]{morton15}
{Morton}, T.~D. 2015, {isochrones: Stellar model grid package}, Astrophysics
  Source Code Library, , , ascl:1503.010

\bibitem[{{Perryman}(2005)}]{perryman05}
{Perryman}, M.~A.~C. 2005, in Astronomical Society of the Pacific Conference
  Series, Vol. 338, Astrometry in the Age of the Next Generation of Large
  Telescopes, ed. P.~K. {Seidelmann} \& A.~K.~B. {Monet}, 3

\bibitem[{{Petigura}(2015)}]{petigura15b}
{Petigura}, E. 2015, PhD Thesis, University of California, arXiv:1510.03902

\bibitem[{{Petigura} {et~al.}(2015){Petigura}, {Schlieder}, {Crossfield},
  {Howard}, {Deck}, {Ciardi}, {Sinukoff}, {Allers}, {Best}, {Liu}, {Beichman},
  {Isaacson}, {Hansen}, \& {L{\'e}pine}}]{petigura15}
{Petigura}, E.~A., {Schlieder}, J.~E., {Crossfield}, I.~J.~M., {et~al.} 2015,
  \apj, 811, 102

\bibitem[{{Pinsonneault} {et~al.}(2012){Pinsonneault}, {An},
  {Molenda-{\.Z}akowicz}, {Chaplin}, {Metcalfe}, \& {Bruntt}}]{pinsonneault11}
{Pinsonneault}, M.~H., {An}, D., {Molenda-{\.Z}akowicz}, J., {et~al.} 2012,
  \apjs, 199, 30

\bibitem[{{Pinsonneault} {et~al.}(2014){Pinsonneault}, {Elsworth}, {Epstein},
  {Hekker}, {M{\'e}sz{\'a}ros}, {Chaplin}, {Johnson}, {Garc{\'{\i}}a},
  {Holtzman}, {Mathur}, {Garc{\'{\i}}a P{\'e}rez}, {Silva Aguirre}, {Girardi},
  {Basu}, {Shetrone}, {Stello}, {Allende Prieto}, {An}, {Beck}, {Beers},
  {Bizyaev}, {Bloemen}, {Bovy}, {Cunha}, {De Ridder}, {Frinchaboy},
  {Garc{\'{\i}}a-Hern{\'a}ndez}, {Gilliland}, {Harding}, {Hearty}, {Huber},
  {Ivans}, {Kallinger}, {Majewski}, {Metcalfe}, {Miglio}, {Mosser}, {Muna},
  {Nidever}, {Schneider}, {Serenelli}, {Smith}, {Tayar}, {Zamora}, \&
  {Zasowski}}]{apokasc}
{Pinsonneault}, M.~H., {Elsworth}, Y., {Epstein}, C., {et~al.} 2014, \apjs,
  215, 19

\bibitem[{{Robin} {et~al.}(2003){Robin}, {Reyl{\'e}}, {Derri{\`e}re}, \&
  {Picaud}}]{robin03}
{Robin}, A.~C., {Reyl{\'e}}, C., {Derri{\`e}re}, S., \& {Picaud}, S. 2003,
  \aap, 409, 523

\bibitem[{{Sanchis-Ojeda} {et~al.}(2015){Sanchis-Ojeda}, {Rappaport},
  {Pall{\`e}}, {Delrez}, {DeVore}, {Gandolfi}, {Fukui}, {Ribas}, {Stassun},
  {Albrecht}, {Dai}, {Gaidos}, {Gillon}, {Hirano}, {Holman}, {Howard},
  {Isaacson}, {Jehin}, {Kuzuhara}, {Mann}, {Marcy}, {Miles-P{\'a}ez},
  {Monta{\~n}{\'e}s-Rodr{\'{\i}}guez}, {Murgas}, {Narita}, {Nowak}, {Onitsuka},
  {Paegert}, {Van Eylen}, {Winn}, \& {Yu}}]{sanchis15}
{Sanchis-Ojeda}, R., {Rappaport}, S., {Pall{\`e}}, E., {et~al.} 2015, \apj,
  812, 112

\bibitem[{{Schlegel} {et~al.}(1998){Schlegel}, {Finkbeiner}, \&
  {Davis}}]{schlegel98}
{Schlegel}, D.~J., {Finkbeiner}, D.~P., \& {Davis}, M. 1998, \apj, 500, 525

\bibitem[{{Schlieder} {et~al.}(2016){Schlieder}, {Crossfield}, {Petigura},
  {Howard}, {Aller}, {Sinukoff}, {Isaacson}, {Fulton}, {Ciardi}, {Bonnefoy},
  {Ziegler}, {Morton}, {L{\'e}pine}, {Obermeier}, {Liu}, {Bailey}, {Baranec},
  {Beichman}, {Defr{\`e}re}, {Henning}, {Hinz}, {Law}, {Riddle}, \&
  {Skemer}}]{schlieder16}
{Schlieder}, J.~E., {Crossfield}, I.~J.~M., {Petigura}, E.~A., {et~al.} 2016,
  \apj, 818, 87

\bibitem[{{Serenelli} {et~al.}(2013){Serenelli}, {Bergemann}, {Ruchti}, \&
  {Casagrande}}]{serenelli13}
{Serenelli}, A.~M., {Bergemann}, M., {Ruchti}, G., \& {Casagrande}, L. 2013,
  \mnras, 429, 3645

\bibitem[{{Sharma} {et~al.}(2011){Sharma}, {Bland-Hawthorn}, {Johnston}, \&
  {Binney}}]{sharma11}
{Sharma}, S., {Bland-Hawthorn}, J., {Johnston}, K.~V., \& {Binney}, J. 2011,
  \apj, 730, 3

\bibitem[{{Sharma} {et~al.}(2014){Sharma}, {Bland-Hawthorn}, {Binney},
  {Freeman}, {Steinmetz}, {Boeche}, {Bienaym{\'e}}, {Gibson}, {Gilmore},
  {Grebel}, {Helmi}, {Kordopatis}, {Munari}, {Navarro}, {Parker}, {Reid},
  {Seabroke}, {Siebert}, {Watson}, {Williams}, {Wyse}, \& {Zwitter}}]{sharma14}
{Sharma}, S., {Bland-Hawthorn}, J., {Binney}, J., {et~al.} 2014, \apj, 793, 51

\bibitem[{{Sinukoff} {et~al.}(2015){Sinukoff}, {Howard}, {Petigura},
  {Schlieder}, {Crossfield}, {Ciardi}, {Fulton}, {Isaacson}, {Aller},
  {Baranec}, {Beichman}, {Hansen}, {Knutson}, {Law}, {Liu}, \&
  {Riddle}}]{sinukoff15}
{Sinukoff}, E., {Howard}, A.~W., {Petigura}, E.~A., {et~al.} 2015, \apj,
  submitted, arXiv:1511.09213

\bibitem[{{Skrutskie} {et~al.}(2006){Skrutskie}, {Cutri}, {Stiening},
  {Weinberg}, {Schneider}, {Carpenter}, {Beichman}, {Capps}, {Chester},
  {Elias}, {Huchra}, {Liebert}, {Lonsdale}, {Monet}, {Price}, {Seitzer},
  {Jarrett}, {Kirkpatrick}, {Gizis}, {Howard}, {Evans}, {Fowler}, {Fullmer},
  {Hurt}, {Light}, {Kopan}, {Marsh}, {McCallon}, {Tam}, {Van Dyk}, \&
  {Wheelock}}]{skrutskie06}
{Skrutskie}, M.~F., {Cutri}, R.~M., {Stiening}, R., {et~al.} 2006, \aj, 131,
  1163

\bibitem[{{Stassun} {et~al.}(2014){Stassun}, {Pepper}, {Paegert}, {De Lee}, \&
  {Sanchis-Ojeda}}]{stassun14}
{Stassun}, K.~G., {Pepper}, J.~A., {Paegert}, M., {De Lee}, N., \&
  {Sanchis-Ojeda}, R. 2014, ArXiv e-prints, arXiv:1410.6379

\bibitem[{{Stello} {et~al.}(2015){Stello}, {Huber}, {Sharma}, {Johnson},
  {Lund}, {Handberg}, {Buzasi}, {Silva Aguirre}, {Chaplin}, {Miglio},
  {Pinsonneault}, {Basu}, {Bedding}, {Bland-Hawthorn}, {Casagrande}, {Davies},
  {Elsworth}, {Garcia}, {Mathur}, {Di Mauro}, {Mosser}, {Schneider},
  {Serenelli}, \& {Valentini}}]{stello15}
{Stello}, D., {Huber}, D., {Sharma}, S., {et~al.} 2015, \apjl, 809, L3

\bibitem[{{Thompson} \& {Fraquelli}(2012)}]{archivemanual}
{Thompson}, S.~E., \& {Fraquelli}, D. 2012, Kepler Archive Manual
  (KDMC-10008-003), http://archive.stsci.edu/kepler/documents.html

\bibitem[{{van Leeuwen}(2007)}]{vanleeuwen07b}
{van Leeuwen}, F. 2007, \aap, 474, 653

\bibitem[{{Vanderburg} \& {Johnson}(2014)}]{vanderburg14}
{Vanderburg}, A., \& {Johnson}, J.~A. 2014, \pasp, 126, 948

\bibitem[{{Vanderburg} {et~al.}(2015){Vanderburg}, {Montet}, {Johnson},
  {Buchhave}, {Zeng}, {Pepe}, {Collier Cameron}, {Latham}, {Molinari}, {Udry},
  {Lovis}, {Matthews}, {Cameron}, {Law}, {Bowler}, {Angus}, {Baranec},
  {Bieryla}, {Boschin}, {Charbonneau}, {Cosentino}, {Dumusque}, {Figueira},
  {Guenther}, {Harutyunyan}, {Hellier}, {Kuschnig}, {Lopez-Morales}, {Mayor},
  {Micela}, {Moffat}, {Pedani}, {Phillips}, {Piotto}, {Pollacco}, {Queloz},
  {Rice}, {Riddle}, {Rowe}, {Rucinski}, {Sasselov}, {S{\'e}gransan},
  {Sozzetti}, {Szentgyorgyi}, {Watson}, \& {Weiss}}]{vanderburg15}
{Vanderburg}, A., {Montet}, B.~T., {Johnson}, J.~A., {et~al.} 2015, \apj, 800,
  59

\bibitem[{{Vanderburg} {et~al.}(2016){Vanderburg}, {Latham}, {Buchhave},
  {Bieryla}, {Berlind}, {Calkins}, {Esquerdo}, {Welsh}, \&
  {Johnson}}]{vanderburg15b}
{Vanderburg}, A., {Latham}, D.~W., {Buchhave}, L.~A., {et~al.} 2016, \apjs,
  222, 14

\bibitem[{{Verner} {et~al.}(2011){Verner}, {Elsworth}, {Chaplin}, {Campante},
  {Corsaro}, {Gaulme}, {Hekker}, {Huber}, {Karoff}, {Mathur}, {Mosser},
  {Appourchaux}, {Ballot}, {Bedding}, {Bonanno}, {Broomhall}, {Garc{\'{\i}}a},
  {Handberg}, {New}, {Stello}, {R{\'e}gulo}, {Roxburgh}, {Salabert}, {White},
  {Caldwell}, {Christiansen}, \& {Fanelli}}]{verner11}
{Verner}, G.~A., {Elsworth}, Y., {Chaplin}, W.~J., {et~al.} 2011, \mnras, 415,
  3539

\bibitem[{{Youdin}(2011)}]{youdin11}
{Youdin}, A.~N. 2011, \apj, 742, 38

\bibitem[{{Zacharias} {et~al.}(2013){Zacharias}, {Finch}, {Girard}, {Henden},
  {Bartlett}, {Monet}, \& {Zacharias}}]{zacharias13}
{Zacharias}, N., {Finch}, C.~T., {Girard}, T.~M., {et~al.} 2013, \aj, 145, 44

\bibitem[{{Zacharias} {et~al.}(2005){Zacharias}, {Monet}, {Levine}, {Urban},
  {Gaume}, \& {Wycoff}}]{zacharias05}
{Zacharias}, N., {Monet}, D.~G., {Levine}, S.~E., {et~al.} 2005, VizieR Online
  Data Catalog, 1297, 0

\end{thebibliography}
